\documentclass[twocolumn]{aastex631}

\usepackage{mathrsfs}
\usepackage{amsmath}

\usepackage[caption=false]{subfig}
\usepackage{nicefrac}

\shorttitle{Searching for changing-state AGNs in massive datasets -- I}
\shortauthors{S\'anchez-S\'aez et al.}
\graphicspath{{./}{figures/}}

\begin{document}

\title{Searching for changing-state AGNs in massive datasets -- I: applying deep learning and anomaly detection techniques to find AGNs with anomalous variability behaviours}

\correspondingauthor{P. S\'anchez-S\'aez}
\email{pasanchezsaez@gmail.com}

\author[0000-0003-0820-4692]{P. S\'anchez-S\'aez}
\affiliation{Inria Chile Research Center, Av. Apoquindo 2827, Las Condes, Chile}
\affiliation{Millennium Institute of Astrophysics (MAS), Nuncio Monse{\~{n}}or S{\'{o}}tero Sanz 100, Providencia, Santiago, Chile} 
\affiliation{Instituto de Astrof{\'{\i}}sica, Facultad de F{\'{i}}sica, Pontificia Universidad Cat{\'{o}}lica de Chile, Casilla 306, Santiago 22, Chile}
\author[0000-0002-5155-587X]{H. Lira}
\affiliation{Inria Chile Research Center, Av. Apoquindo 2827, Las Condes, Chile}
\author[0000-0003-2789-5062]{L. Mart\'i}
\affiliation{Inria Chile Research Center, Av. Apoquindo 2827, Las Condes, Chile}
\author[0000-0002-5037-9974]{N. S\'anchez-Pi}
\affiliation{Inria Chile Research Center, Av. Apoquindo 2827, Las Condes, Chile}
\author[0000-0002-2045-7134]{J. Arredondo}
\affiliation{Millennium Institute of Astrophysics (MAS), Nuncio Monse{\~{n}}or S{\'{o}}tero Sanz 100, Providencia, Santiago, Chile} 
\author[0000-0002-8686-8737]{F. E. Bauer}
\affiliation{Instituto de Astrof{\'{\i}}sica, Facultad de F{\'{i}}sica, Pontificia Universidad Cat{\'{o}}lica de Chile, Casilla 306, Santiago 22, Chile} 
\affiliation{Centro de Astroingenier{\'{\i}}a, Pontificia Universidad Cat{\'{o}}lica de Chile, Av. Vicu\~{n}a Mackenna 4860, 7820436 Macul, Santiago, Chile} 
\affiliation{Millennium Institute of Astrophysics (MAS), Nuncio Monse{\~{n}}or S{\'{o}}tero Sanz 100, Providencia, Santiago, Chile}
\affiliation{Space Science Institute, 4750 Walnut Street, Suite 205, Boulder, Colorado 80301} 
\author[0000-0001-7868-7031]{A. Bayo}
\affiliation{Instituto de F\'isica y Astronom\'ia, Facultad de Ciencias, Universidad de Valpara\'iso, Av. Gran Breta\~na 1111, Valpara\'iso, Chile}
\affiliation{N\'ucleo Milenio de Formaci\'on Planetaria (NPF), Valpara\'iso, Chile}
\author[0000-0002-2720-7218]{G. Cabrera-Vives}
\affiliation{Department of Computer Science, Universidad de Concepci\'on, Edmundo Larenas 219, Concepci\'on, Chile}
\affiliation{Millennium Institute of Astrophysics (MAS), Nuncio Monse{\~{n}}or S{\'{o}}tero Sanz 100, Providencia, Santiago, Chile} 
\author{C. Donoso-Oliva}
\affiliation{Department of Computer Science, Universidad de Concepci\'on, Edmundo Larenas 219, Concepci\'on, Chile}
\affiliation{Millennium Institute of Astrophysics (MAS), Nuncio Monse{\~{n}}or S{\'{o}}tero Sanz 100, Providencia, Santiago, Chile} 
\author[0000-0001-9164-4722]{P. A. Est\'evez}
\affiliation{Millennium Institute of Astrophysics (MAS), Nuncio Monse{\~{n}}or S{\'{o}}tero Sanz 100, Providencia, Santiago, Chile} 
\affiliation{Department of Electrical Engineering, Universidad de Chile, Av. Tupper 2007, Santiago 8320000, Chile}
\author[0000-0003-4723-9660]{S. Eyheramendy}
\affiliation{Faculty of Engineering and Sciences, Universidad Adolfo Iba\~nez, Diagonal Las Torres 2700, Pe\~nalol\'en, Santiago, Chile}
\affiliation{Millennium Institute of Astrophysics (MAS), Nuncio Monse{\~{n}}or S{\'{o}}tero Sanz 100, Providencia, Santiago, Chile}
\author[0000-0003-3459-2270]{F. F\"orster}
\affiliation{Data and Artificial Intelligence Initiative (D\&IA), University of Chile.}
\affiliation{Millennium Institute of Astrophysics (MAS), Nuncio Monse{\~{n}}or S{\'{o}}tero Sanz 100, Providencia, Santiago, Chile} 
\affiliation{Center for Mathematical Modeling, University of Chile, AFB170001, Chile}
\affiliation{Departamento de Astronom\'ia, Universidad de Chile, Casilla 36D, Santiago, Chile}
\author[0000-0002-8606-6961]{L.~Hern\'andez-Garc\'ia}
\affiliation{Millennium Institute of Astrophysics (MAS), Nuncio Monse{\~{n}}or S{\'{o}}tero Sanz 100, Providencia, Santiago, Chile} 
\affiliation{Instituto de F\'isica y Astronom\'ia, Facultad de Ciencias, Universidad de Valpara\'iso, Av. Gran Breta\~na 1111, Valpara\'iso, Chile}
\author[0000-0002-8722-516X]{A. M. Mu\~noz Arancibia}
\affiliation{Millennium Institute of Astrophysics (MAS), Nuncio Monse{\~{n}}or S{\'{o}}tero Sanz 100, Providencia, Santiago, Chile}
\affiliation{Center for Mathematical Modeling, University of Chile, AFB170001, Chile}
\author[0000-0003-4644-8698]{M. P\'erez-Carrasco}
\affiliation{Department of Computer Science, Universidad de Concepci\'on, Edmundo Larenas 219, Concepci\'on, Chile}
\affiliation{Millennium Institute of Astrophysics (MAS), Nuncio Monse{\~{n}}or S{\'{o}}tero Sanz 100, Providencia, Santiago, Chile} 
\author{M. Sep\'ulveda}
\affiliation{Departamento de Astronom\'ia, Universidad de Chile, Casilla 36D, Santiago, Chile}
\author[0000-0001-6699-4181]{J. R. Vergara}
\affiliation{Department of Computing, Universidad Tecnol\'{o}gica Metropolitana, Santiago, Chile}
\affiliation{Millennium Institute of Astrophysics (MAS), Nuncio Monse{\~{n}}or S{\'{o}}tero Sanz 100, Providencia, Santiago, Chile} 




\begin{abstract}
The classic classification scheme for Active Galactic Nuclei (AGNs) was recently challenged by the discovery of the so-called changing-state (changing-look) AGNs (CSAGNs). The physical mechanism behind this phenomenon is still a matter of open debate and the samples are too small and of serendipitous nature to provide robust answers. In order to tackle this problem, we need to design methods that are able to detect AGN right in the act of changing–state. Here we present an anomaly detection (AD) technique designed to identify AGN light curves with anomalous behaviors in massive datasets. The main aim of this technique is to identify CSAGN at different stages of the transition, but it can also be used for more general purposes, such as cleaning massive datasets for AGN variability analyses. We used light curves from the Zwicky Transient Facility data release 5 (ZTF DR5), containing a sample of 230,451 AGNs of different classes. The ZTF DR5 light curves were modeled with a Variational Recurrent Autoencoder (VRAE) architecture, that allowed us to obtain a set of attributes from the VRAE latent space that describes the general behaviour of our sample. These attributes were then used as features for an Isolation Forest (IF) algorithm, that is an anomaly detector for a ``one class'' kind of problem. We used the VRAE reconstruction errors and the IF anomaly score to select a sample of 8,809 anomalies. These anomalies are dominated by bogus candidates, but we were able to identify 75 promising CSAGN candidates. 
\end{abstract}




\section{Introduction} \label{sec:intro}

Active Galactic Nuclei (AGNs) are powered by the release of gravitational energy related with the accretion of material onto a supermassive black hole (SMBH; \citealt{Lynden-Bell69}). They are characterized by their time-variable emission, and these variations can be observed from the gamma-rays to the radio wavebands. The light curves of AGNs appear to be stochastic and have characteristic timescales that range from hours to years (e.\,g., \citealt{Ulrich97,Padovani17}).


Changing-state AGNs (CSAGNs; sometimes called optical changing-look AGNs), correspond to AGNs that change their classification as type 1 [objects which present broad permitted emission lines ($\text{FWHM} \gtrsim 2000$ km s$^{-1}$) in their spectra] or type 2 AGN (objects that lack broad emission lines in their spectra), as well as to sources that present large changes in the flux of their broad emission lines. These changes are normally observed on timescales between months and years (e.\,g., \citealt{LaMassa15, MacLeod16, Stern18, Ross18, MacLeod19, Graham20}). This transition phase can be accompanied by a drastic change in the AGN continuum flux, by orders of magnitudes which is unexpected considering the typical variability amplitudes observed in large AGN samples ($\sim$0.1--0.2 magnitudes per year; e.\,g., \citealt{VandenBerk04,Sanchez-Saez18}).

The origin of the anomalous optical variability of CSAGNs is still not well understood. Possible physical explanations include changes in obscuration (which produce variable absorption), or changes in the black hole accretion rate or accretion disk structure. Recent observations indicate that this phenomenon (in the optical range) is not due to obscuration; since this would imply large cloud size and rapid velocities required to occult a sizable portion of disk; but is rather due to changes in the accretion disk innermost regions (e.\,g. \citealt{LaMassa15,Stern18,Ross18,MacLeod19}). Understanding the changing--state phenomenon is therefore crucial to improve our knowledge of the physics behind AGN variability. In particular, we need to understand to what extent the variability properties of CSAGN are generally outliers from the distribution of AGN overall (i.e., even before they exhibit their strong changes). 

Most of the few known CSAGNs have been discovered ``after the fact'', using archival data, not allowing the possibility to witness the change / follow it up, in a multi-wavelength manner, and thus, innovative strategies are needed to catch them during the transition phase. \cite{Ricci20} presented observations of the disappearance and re-appearance of the X-ray corona in 1ES 1927+654, a source that was previously classified as a CSAGN by \cite{Trakhtenbrot19}. This event is particularly interesting because it is a CSAGN both in the optical and X-ray regimes. The authors claim that a tidal disruption event (TDE; tidal disruption of a star by the accreting SMBH) provoked an increase in the accretion rate at the inner-most regions of the accretion disk, which emptied the inner disk, and led to the destruction of the X-ray corona. This corresponds to the first CSAGN extensively monitored at X-rays and optical through a transition state, placing strong constraints on the physics of the system. That work demonstrated the importance of the rapid detection and rapid follow-up of changing-state events, which can give us the opportunity to probe accretion physics.

Previous searches of CSAGN in massive archival datasets have employed simple statistical analysis to find them, using for instance measurements of the variability amplitude (e.\,g.; \citealt{MacLeod16,MacLeod19}), or more complex analysis like Bayesian blocks representation (BB; \citealt{Scargle13,Graham20}). However, these methods are not suitable for the real-time detection of CSAGN events, since they favor the detection of events well after they have occurred, and not the detection of the transition phase. More promising techniques involve time series forecasting or the use of Anomaly Detection algorithms (AD; e.\,g., \citealt{Mehrotra17}).

AD techniques have been used in astronomy to search for unusual objects within astronomical datasets, however this search has largely focused to date on transient and periodic classes (e.\,g. \citealt{Ishida19,Lochner20,Malanchev20, Villar21}). P\'erez-Carrasco et al. (in prep) presents the outlier detector for the Automatic Learning for the Rapid Classification of Events (ALeRCE) broker \citep{Forster21, Carrasco-Davis20, Sanchez-Saez21}. Its main goal is to find multiple types of variable and transient sources, that are not included in the taxonomy tree of the ALeRCE broker light curve classifier \citep{Sanchez-Saez21}. However, AD algorithms can also be used to detect light curves of objects with previous known classifications, that suddenly start presenting unusual variability behaviours (also known as contextual AD). Recently, \cite{Suberlak21} showed that CSAGN candidates appear as outliers when they are compared with the damped random walk (DRW) parameters (timescale and variance; \citealt{Kelly09}) obtained from light curves of the same objects, that cover different timespans (and using different datasets). Thus, we can expect that the early detection of CSAGNs would benefit from the use of AD techniques on current large monitoring surveys.

In this work we present an unsupervised AD algorithm to search for AGN light curves with anomalous variability behaviours in massive datasets, focusing in particular on data from the Zwicky Transient Facility (ZTF; \citealt{Bellm19}). The main goal of the proposed algorithm is to find CSAGN candidates at different stages of the changing-state event (i.e., either early or late stages), although the methodology should be generally useful to find other classes of anomalous light curves, like 
atypical flaring activity (e.\,g., \citealt{Graham17,Trakhtenbrot19NatAs,Frederick20}), extremely variable AGNs (e.\,g., \citealt{Rumbaugh18,Guo20,Luo20}), or sources with incorrect labels or photometric issues, as we show in the following sections. Our method is inspired by the work presented in \cite{Tachibana20}, who used a Recurrent Autoencoder (RAE) architecture to model AGN light curves observed by the Catalina Real-time Transient Survey (CRTS; \citealt{Drake09}). We instead use a Variational Recurrent Autoencoder (VRAE; \citealt{Fabius14}) architecture together with an Isolation Forest (IF; \citealt{Liu08}) algorithm, following a two-stage approach, to search for anomalous light curves. As far as we know, this is the first attempt to search for anomalous AGN variability behaviors using AD techniques.   

The paper is organized as follows. In Section \ref{sec:data} we describe the datasets used for this work, including the AGN catalogs considered, and the procedure to define the final sample of ZTF light curves. In Section \ref{sec:methods} we describe the VRAE architecture, and the AD methodology used. In Section \ref{sec:results} we present a comparison of the results obtained using RAE and VRAE architectures. We show the results obtained using our two-stage AD method, together with a description of the anomalies found. In Section \ref{sec:discussion} we discuss our main findings and compare our results with previous works. Finally, in Section \ref{sec:summary} we summarize and conclude the main results obtained in this work. 

\section{Data} \label{sec:data}

\subsection{The sample of known AGNs}\label{subsec:catalogs}

In order to detect CSAGNs in real-time, we need a sample of AGNs with known classifications. In particular, we need sources optically classified as type 1 or type 2. However, in order to test the quality of our AD algorithm, we decided to include other classes of AGNs, such as blazars, X-ray detected, and radio detected AGNs. We used the following AGN catalogs to create our sample of known AGNs:

\begin{itemize}
    \item The Roma-BZCAT Multi-Frequency Catalog of Blazars (ROMABZCAT; \citealt{Massaro15}; priority 1).
    \item The New Catalog of Type 1 AGNs (Oh2015; \citealt{Oh15}; priority 2).
    \item The Million Quasars (MILLIQUAS; priority 3) Catalog, Version 7.0 \citep{Flesch15,Flesch19}. 
\end{itemize}

These catalogs were used to obtain celestial coordinates and classifications of known AGNs. We considered the following 12 classes (labels used in this work are provided in parenthesis):
\begin{itemize}
    \item Seyfert 1 or host-dominated type 1 AGN (A).
    \item Seyfert 1 with radio emission (AR).
    \item QSO 1 or core-dominated type 1 AGN (Q).
    \item QSO 1 with radio emission (QR).
    \item Host- and core-dominated type 2 AGN (type2).
    \item Type 2 AGN with radio emission (type2 R).
    \item BL Lac AGN (BLLac).
    \item Flat Spectrum Radio Quasar (FSRQ).
    \item Blazars of Uncertain type (BZU).
    \item AGN identified by its X-ray emission (X).
    \item AGN identified by its radio emission (R).
\end{itemize}
If a source appeared in more than one catalog, we followed the priorities defined above to select the final label. For instance, if a source appears in ROMABZCAT and in MILLIQUAS we kept the classification provided by ROMABZCAT. After this, we ended up with a sample of 875,322 known AGNs (sources classified as candidates in MILLIQUAS were not included).

In addition, we obtained physical properties (redshifts, SMBH masses, Eddington ratios and luminosities) from Sloan Digital Sky Survey (SDSS; \citealt{York00}) for 524,454 sources in the sample of known AGNs, using the Oh2015 catalog, and the catalog of Spectral Properties of Quasars from SDSS Data Release 14 \citep{Rakshit20}. This information was used to define training and validation sets balanced by means of their physical properties (see Section \ref{subsec:traningsets}).

\subsection{Optical light curves}\label{subsec:lcs}

For this work we used the point spread function (PSF) fit-based optical light curves available in the ZTF public data release 5\footnote{ Documentation: \url{https://www.ztf.caltech.edu/page/dr5}} (ZTF DR5; \citealt{Masci19}). In particular, we used the ZTF Application Programming Interface (ZTF API) to obtain light curves for all the sources in our sample of known AGNs described in Section \ref{subsec:catalogs}. We retrieved the light curves of each source, centred in the coordinates of the catalogs presented in Section \ref{subsec:catalogs}, and using a radius of 1.5 arcseconds, which is the standard cross-matching radius used by ZTF. When doing this for a particular object, the ZTF API can return more than one light curve in a given filter. This happens because ZTF treats independently the light curves observed in a particular field, filter, and CCD-quadrant, and thus, if a source appears in more than one combination of these three, ZTF will provide more than one light curve. The ZTF quadrants are calibrated independently, and thus combining light curves from different fields and CCD-quadrants (but same filter), can produce spurious variability \citep{vanRoestel21}. Therefore, to avoid having multiple light curves for any particular source, we only used the longest light curve (in terms of its timespan) of each filter of a particular object. 

ZTF DR5 provides light curves in the $g$, $r$ and $i$ ZTF bands, however for this work we only used light curves observed in the $g$ band. We expect that this band presents less contamination from the host galaxy of each target. Future work will further exploit the potential of using multi-band light curves for AD.  

We cleaned the ZTF DR5 light curves using the \texttt{catflags} quality score, keeping only the epochs with \texttt{catflags}=0, as advised by the ZTF documentation. We also removed all the epochs with a photometric error higher than 1 magnitude.  

The ZTF PSF-fit-based light curves are provided for point-like and extended sources, however PSF-fit photometry measurements are proper for point-sources only. Therefore, we used the catalog presented in \cite{Tachibana18}, which provides a morphological classification score (\texttt{ps\_score}) using PanSTARRS1 photometry, to filter-out extended sources from our analysis, keeping only sources with \texttt{ps\_score}$>0.5$ (i.e., point sources according to \citealt{Tachibana18}).

In addition, to ensure the detection of AGN variability (with timescales of hours to years; \citealt{Ulrich97}), we filtered the sample by the timespan ($t_{\text{length}}$) and number of epochs ($n_{\text{det}}$) of the light curves, keeping only sources whose light curves have $t_{\text{length}}\geq 730 \; \text{days}$ (2 years) and $n_{\text{det}}\geq 50$. We also removed from the sample sources with an average $g$ band magnitude fainter than 20.6 (close to the limiting magnitude of the ZTF images), and brighter than 13.5 (to avoid saturated observations).

Finally, we computed two variability features; $P_{\text{var}}$, the probability that the source is intrinsically variable, and $\sigma^2_{\text{rms}}$, the normalized excess variance (for a detailed description see Appendix A of \citealt{Sanchez-Saez21}); and used them to remove from the sample light curves that do not show any variations. Following a similar approach to \cite{Sanchez17}, we removed from the analysis sources with $P_{\text{var}}<0.95$ and $\sigma^2_{\text{rms}}<0$. After this, we ended up with a sample of 230,451 known AGNs with variable (and long enough) ZTF light curves (hereafter, the ``full'' sample), with 166,243 of these having spectroscopic properties measured from SDSS (hereafter, the ``full spectroscopic'' sample). 

\begin{figure}[tb]
    \centering
    \includegraphics[width=\linewidth]{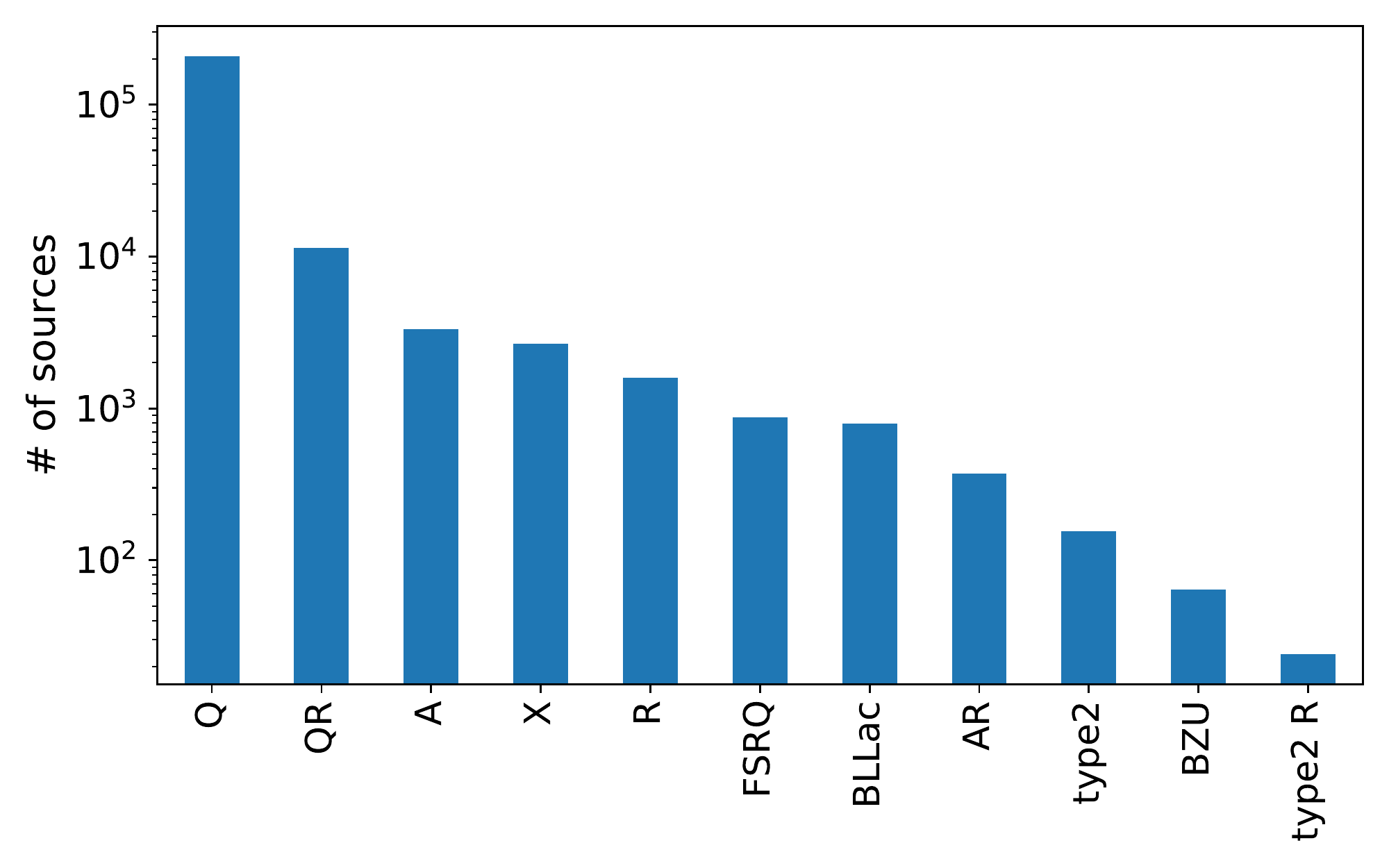}
    \caption{Number of sources per class in the full sample.\label{figure:num_classes_ls}}
\end{figure}

Figure \ref{figure:num_classes_ls} shows the number of sources per class in the full sample. It is dominated by sources of class ``Q'', which correspond to 90.7\% of the total. 

\subsection{Training, validation, and testing sets}\label{subsec:traningsets}

It is well known that AGNs with different physical properties show different variability behaviors (e.\,g., \citealt{MacLeod10,Hernandez-Garcia16,Caplar17,Sanchez-Saez18}). For instance, SMBH mass ($\mathrm{BH}_{\mathrm{mass}}$) and accretion rate set the luminosity variability properties to first order. Distance sets the flux limit, with rare sources only found in large volumes, and faint sources only found in small volumes. Finally, past follow-up and multi-wavelength coverage may include its own set of biases. Thus, in order to develop an AD algorithm that is not biased towards the more common physical properties in the sample, we need to train our algorithm using training and validation sets that are balanced by means of these physical properties. Otherwise, sources with physical properties that escape the general distribution will appear as anomalous, even though they do not show anomalous variability behaviors.

\begin{figure*}[tb]
    \centering
    \includegraphics[width=\linewidth]{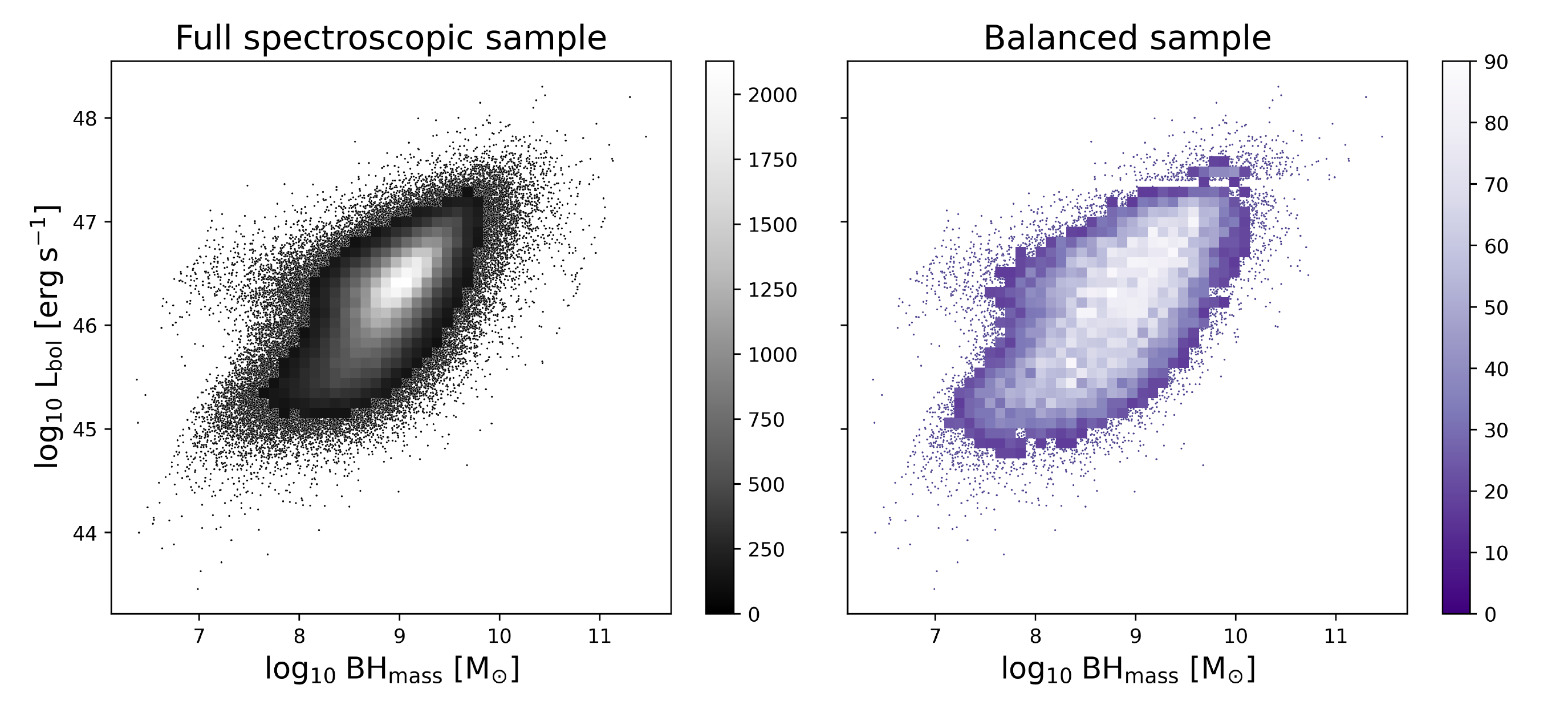}
    \caption{Bivariate histograms of the black hole mass ($\mathrm{BH}_{\mathrm{mass}}$) and the bolometric luminosity ($\mathrm{L}_{\mathrm{bol}}$) of the full spectroscopic sample (left) and the balanced sample (right).\label{figure:spec_prop}}
\end{figure*}

The left panel of Figure \ref{figure:spec_prop} shows $\mathrm{BH}_{\mathrm{mass}}$ and the bolometric luminosity ($\mathrm{L}_{\mathrm{bol}}$) of the full spectroscopic sample (166,243 sources). From the figure we can see that most of the sources cluster around $\mathrm{BH}_{\mathrm{mass}}\sim10^9$ and $\mathrm{L}_{\mathrm{bol}}\sim10^{46.5}$. In order to have homogeneous training and validation sets, we defined two sub-sample (of the full sample with spectral properties) that are balanced by means of their $\mathrm{BH}_{\mathrm{mass}}$ and $\mathrm{L}_{\mathrm{bol}}$.

\begin{figure}[tb]
\centering
\includegraphics[width=\linewidth]{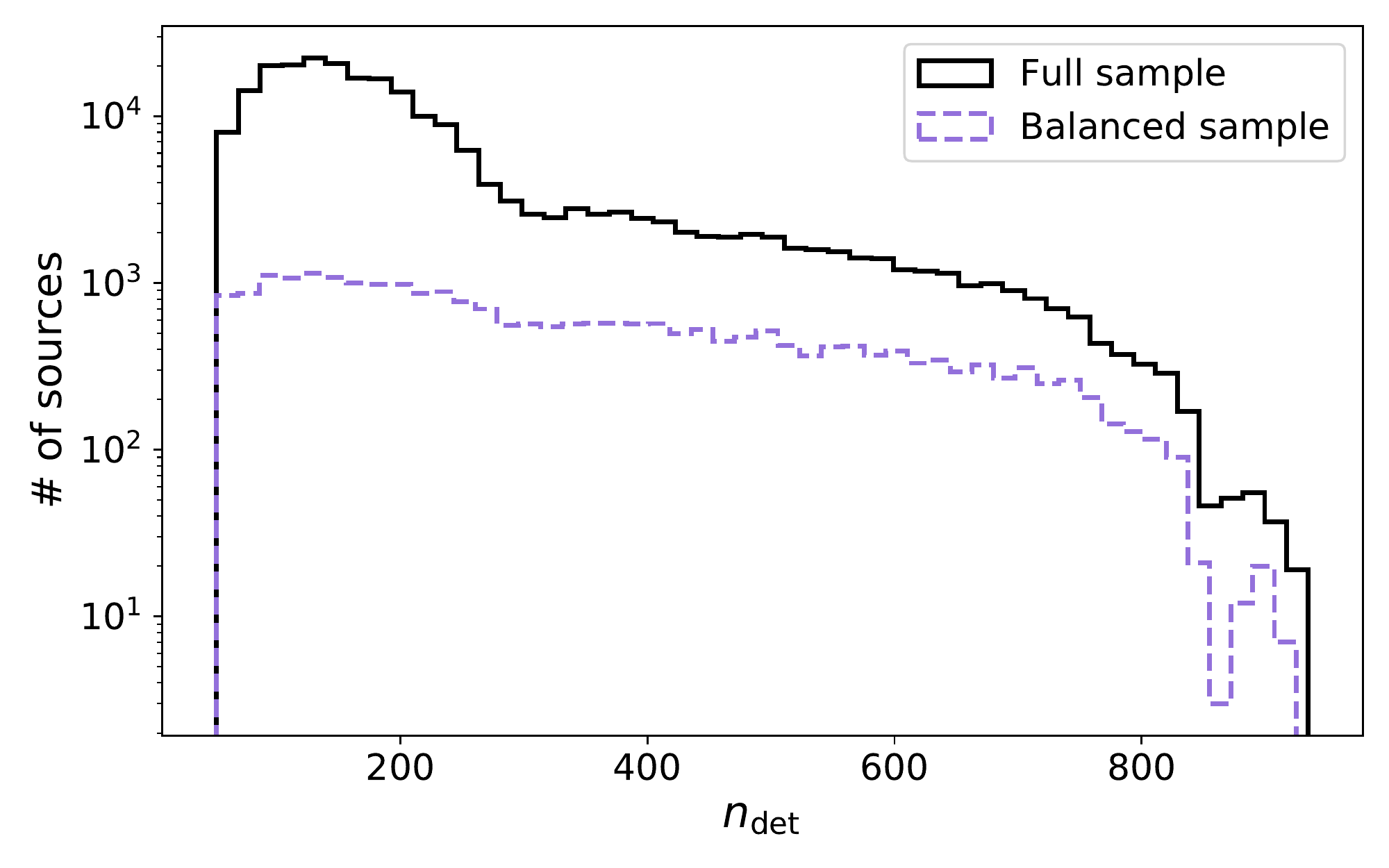}
\caption{Histogram of the number of epochs ($n_{\text{det}}$) per light curve for the full sample (black) and the balanced sample (purple).\label{figure:ndet}}
\end{figure}

In addition, the number of epochs per light curve covers a wide range in our sample, with $50 \leq n_{\text{det}} \leq 935$. Figure \ref{figure:ndet} shows in black the histogram of $n_{\text{det}}$ per source for the full sample. It can be seen that most of the light curves have less than 200 epochs, and only a few have more than 800 epochs. This can produce some issues to our AD algorithm (see discussion in Sections \ref{subsec:vrae} and \ref{subsec:performance}), and therefore, we decided to balance one of the balanced sub-samples by the number of epochs as well. 

Accordingly, we constructed two training and validation sets, one that is balanced only by means of the physical properties (hereafter, the ``balanced phys'' sample), and another one balanced by means of the physical properties and $n_{\text{det}}$ (hereafter, the ``balanced phys-epochs'' sample). 

To construct the balanced phys sample, we defined a grid in the space of $\mathrm{BH}_{\mathrm{mass}}$ and $\mathrm{L}_{\mathrm{bol}}$, were each bin had a width of 0.1 dex, and selected a random sample of 40 targets from each bin (if the bin has less than 40 targets, we kept all the sources). From this, we ended up with a sample of 26,614 objects, with 88\% of them belonging to the class Q. 

On the other hand, to construct balanced phys-epochs sample, we defined a grid in the space of $n_{\text{det}}$, $\mathrm{BH}_{\mathrm{mass}}$, and $\mathrm{L}_{\mathrm{bol}}$, were each bin had a width of 0.1 dex for  $\mathrm{BH}_{\mathrm{mass}}$ and $\mathrm{L}_{\mathrm{bol}}$, and a width of 20 for $n_{\text{det}}$, and selected a random sample of 2 targets from each bin (if the bin has less than 2 targets, we kept all the sources). By doing this, we ended up with a balanced subset of 24,755 sources. According to the labels of the original catalogs, 95\% of this balanced sample is classified as Q. The right panel of Figure \ref{figure:spec_prop} shows the distribution of $\mathrm{BH}_{\mathrm{mass}}$, and $\mathrm{L}_{\mathrm{bol}}$ for the balanced phys-epochs sample, and Figure \ref{figure:ndet} shows in purple its histogram of $n_{\text{det}}$. From both figures we can see that this subset is more balanced in their physical and observational properties, compared to the original full set. For the case of the balanced phys sample, the distributions of the physical properties are similar to those shown in Figure \ref{figure:spec_prop} for the balanced phys-epochs sample.

The balanced phys set is used only for testing purposes, while the balanced phys-epochs set is used to train the final AD model, as explained in the following sections. We used 20\% of each balanced sample as validation set, and 80\% as training set. 

\section{Methods} \label{sec:methods}

In this work we followed an unsupervised approach (i.e., without considering the labels of our dataset) to search for anomalous AGN variability behaviors. We followed a two step approach, where we first train a VRAE architecture to reconstruct the light curves in our sample, and used the reconstruction error as an AD score. Then, we used the VRAE mean latent space as a vector of features for an IF algorithm, and selected anomalies using the anomaly score provided by the IF method. In the following section we provide a detailed description of our AD methodology. 

\subsection{VRAE architecture}\label{subsec:vrae}

Autoencoders (AEs) are artificial neural networks that are able to learn compressed representations of an input dataset, called the ``latent representations'', in an unsupervised way. The latent representations are then used to reconstruct or predict the original data. Basically, AEs are composed by an encoder set of layers, that compress the data, a latent space, that contains the compressed information, and a decoder set of layers that produces a reconstruction or prediction. RAEs, on the other hand, are recurrent AEs that use recurrent neural networks (RNN) to encode and decode sequential data, and are normally used for sequence-to-sequence learning.

\cite{Tachibana20} presented a novel algorithm to empirically model AGN light curves using an RAE (e.\,g., \citealt{Naul18}) architecture. One of the main advantages of using this kind of technique is that it does not assume any underlying statistical process when fitting the light curves (as for instance DRW modelling does; e.\,g., \citealt{Kelly09}), but instead, it follows a nonparametric approach that learns by itself the most important attributes that describe the data. This allows the algorithm to learn properties that might be hard to recognize or characterize by any expert and/or parametrize within a model (statistically or physically motivated).

Instead of using an RAE architecture, we adopt a VRAE architecture to extract a set of latent attributes or features from the AGN light curves, since it has been demonstrated that variational autoencoders (VAEs; \citealt{Kingma14}) are more suitable for AD tasks (e.\,g., \citealt{Portillo20,Villar21}).

\begin{figure*}[tb]
    \centering
    \includegraphics[width=0.92\linewidth]{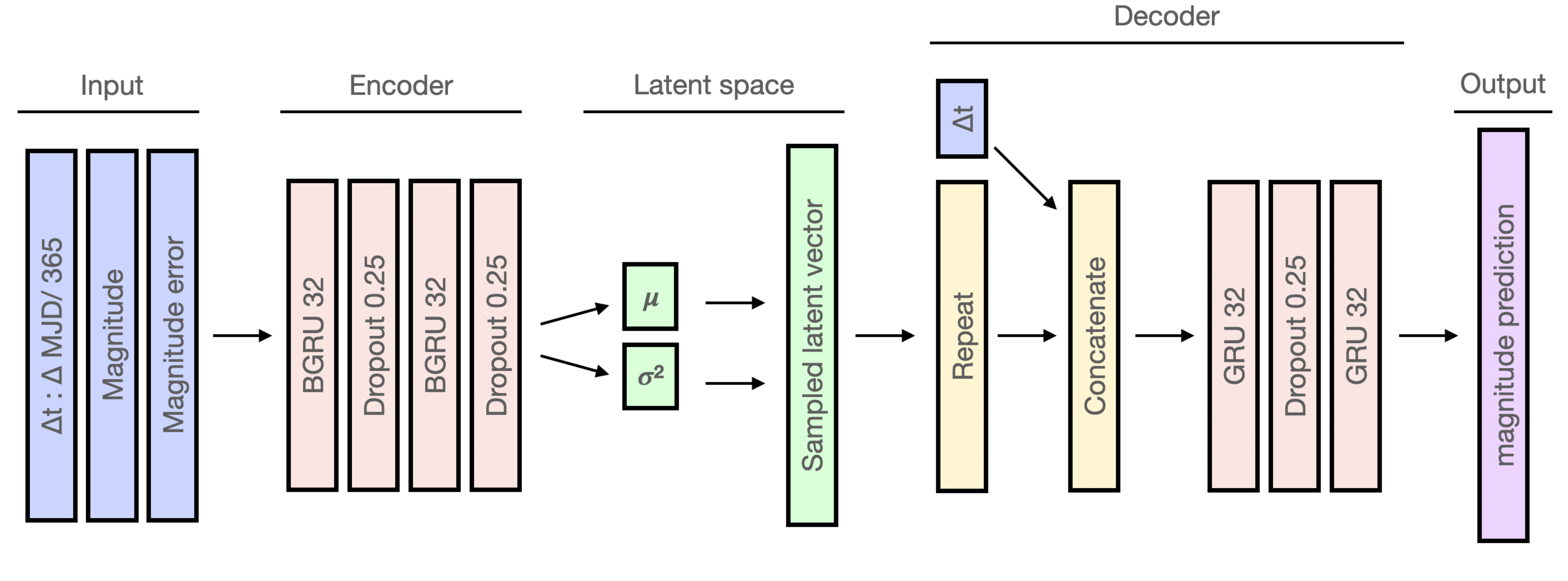}
    \caption{Proposed VRAE architecture. The model receives as input $\Delta t$, normalized magnitude, and the normalized magnitude error, and outputs a prediction of the normalized magnitude for the original $\Delta t$ vector.  \label{figure:architecture}}
\end{figure*}

VAEs\footnote{\citealt{Portillo20} provides an astronomer friendly description of the mathematics behind AEs and VAEs.} correspond to a modification of the more classical AE architectures. In the case of VAE architectures, the latent representations are described by multivariate normal distributions, where each attribute or feature in the latent space is described by a latent mean ($\mu$) and a latent variance ($\sigma^2$), which can be used to randomly sample a set of attributes. Then, we can use this sampled vector of latent attributes to generate a prediction or reconstruction of the input data. Since we can generate multiple latent vectors from a given distribution of $\mu$ and $\sigma^2$, we can use VAEs as generative models, and generate (or sample) for instance new light curves that can be used for data augmentation (e.\,g., \citealt{Moreno-Barea20}). By constructing the latent space in this way, we are also forcing the latent space to be smooth and to cluster together subsets with similar reconstructions (e.\,g., light curves with similar shapes or properties are closer in the latent space), and thus facilitating the detection of anomalous data. 

VRAEs \citep{Fabius14} are RNNs based on VAEs that encode sequential data into a variational distribution over latent variables. One of the main advantages of using RNNs is that they can deal with time series of different lengths, and thus are ideal for light curve analyses. \cite{Tachibana20} used long short-term memory (LSTM; \citealt{Hochreiter97}) layers, while we use gated recurrent units (GRU; \citealt{Cho14}) neurons instead, as they provide similar results and are faster to train \citep{Yang20}. 

Figure \ref{figure:architecture} shows the VRAE architecture used in this work. It is similar to the one used by \cite{Tachibana20}, but, as already mentioned, instead of an AE we used a VAE, and instead of LSTM units we used GRUs. In order to deal with the unevenly sampled ZTF light curves, we used the following input values:
\begin{itemize}
    \item $\Delta t$: differences between sampling times, normalized by 365 days (to use units of years).
    \item Normalized magnitude ($m_{norm}$): magnitude normalized by the average magnitude and the standard deviation of the light curve [$m_{norm} =(m-\overline{m})/\text{std}(m)$].
    \item Normalized magnitude errors (err$_{norm}$): photometric errors (err) normalized by the standard deviation of the light curve [err$_{norm}=$err/$\text{std}(m)$]
\end{itemize}

We used two Bidirectional GRU layers (BGRU\footnote{Where a sequence is passed through a forward GRU and through a backwards GRU, and the outputs of the two are used to compute the output of the layer (e.\,g., \citealt{Chaini20}).}) for the encoder and two GRU layers for the decoder, each one with 32 units, with dropout of 0.25, and with linear activation functions. The latent space mean and variance correspond to Dense layers (regular fully connected neural networks) of size 16, from which a sampled latent vector of size 16 is randomly generated. In the initial layers of the decoder, the sampled latent vector is repeated (in a copy-paste manner) $N_{T}$ times, with $N_{T}$ being the number of epochs in the light curves, which in this case we define as the maximum number of $\Delta t$ values in the given batch being evaluated (934 when using the full sample at once), and the values of each $\Delta t$ are concatenated to this repeated vector (for shorter light curves the missing $\Delta t$ values are replaced by zero). This is done to allow the interpolation and/or extrapolation of the light curves if needed.\footnote{This particular property of our architecture is not used in this work, but it will be used in future analyses, and will allow us to do, for instance, light curve forecasting.} This concatenated matrix is then the input of the decoder GRU layers. Finally, we end up with a vector of predicted (or reconstructed) normalized magnitudes for each input light curve ($m_{rec}$). 

VAEs in general are trained using a loss function that includes two terms, one that takes into account the reconstruction error, called reconstruction loss, and another one, called latent loss, that forces the latent space to look as if it was sampled from a prior distribution, which in this case we assume to be a Normal distribution. Thus, for the reconstruction loss we used the weighted mean squared error (MSE), where the weights correspond to the inverse of the normalized photometric errors ($w=\nicefrac{1}{\text{err}_{norm}}$), and for the case of the latent loss, we used the standard Kullback–Leibler (KL) divergence:
\begin{multline}
\mathscr{L}=\frac{1}{N_{lc}}\sum_k^{N_{lc}}\sum_i^{N_T}\left[\left(m_{{norm}_i} - m_{rec_{i}}\right) w_i \right]^2\\
-\frac{1}{200}\frac{1}{N_{lc}}\sum_k^{N_{lc}}\sum_{j}^{N_l}0.5 \left[1+\log\left(\sigma^2_j\right) - \sigma^2_j - \mu^2_j \right]\,
\end{multline}
with $N_{lc}$ being the number of light curves in the batch, and $N_l$ the dimensionality of the latent space. Note that we divide the latent loss by 200. This is done to ensure that the latent and reconstruction losses have similar scales, since when we compute the reconstruction loss, we are computing the mean over the total number of epochs of each light curve. We selected 200 because it is a representative number of the total number of epochs of the light curves. This number can be tuned as a hyper-parameter (e.\,g., $\beta$-VAE; \citealt{Higgins17}), but we leave that for future work, as good results are achieved by assuming this particular value. 

The model was trained in \texttt{Google Colaboratory} \citep{Bisong19}, taking advantage of the graphics processing units (GPUs) offered by its serverless Jupyter notebook environment. We used \texttt{Keras} \citep{Keras} with a \texttt{TensorFlow} backend \citep{Tensorflow}. We minimize the loss function using the Adam optimizer \citep{Kingma14}, with standard parameter values ($\beta_1=0.9$ and $\beta_2=0.999$), a learning rate of $10^{-3}$, and a batch size of 512. We tested modifications of the original proposed architecture (e.\,g., different number of units or layers), but we did not observe any relevant improvements in the results, thus we decided to keep the original number of units and layers. Finally, to train our architecture, we used the balanced sets described in Section \ref{subsec:traningsets}.

\subsection{AD methodology}\label{subsec:ad}

A common way to do time series AD with RAEs or VRAEs is to train a model with a training set that is assumed to be normal, and then use this model to reconstruct the data of a test set and measure the reconstruction error. If the reconstruction error is above a given threshold, then the time series is labeled as anomalous (e.\,g., \citealt{Malhotra16,Pereira18}).

\begin{figure}[tb]
\centering
\includegraphics[width=\linewidth]{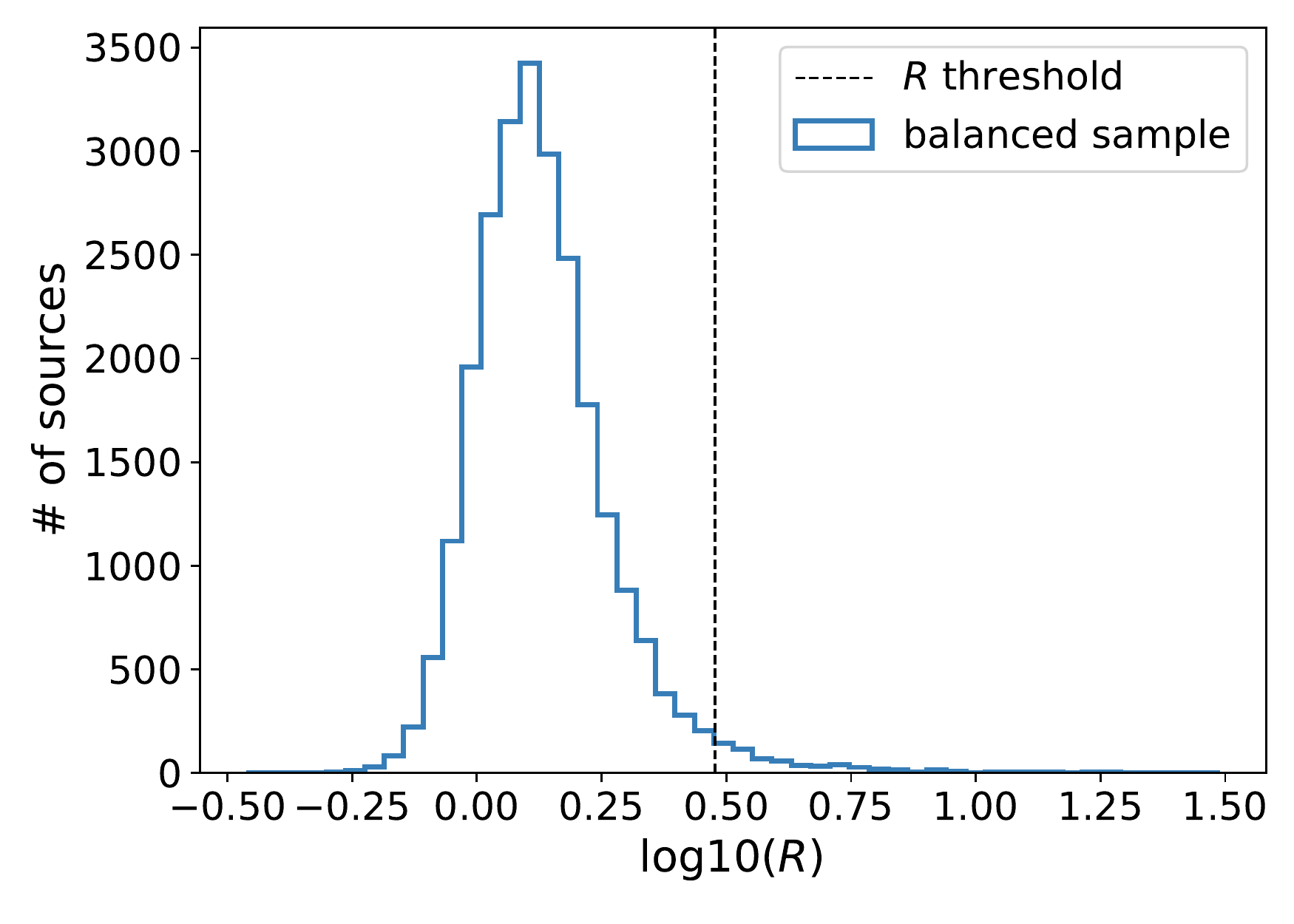}
\caption{Histogram of the reconstruction error ($R$) per light curve for the balanced phys-epochs set. The threshold of $R=3$ is shown with a black dashed line.}
\label{figure:R_balsample}
\end{figure}

Another option, is to use the RAE or VRAE latent space as features for an AD algorithm, such as IF. IF is a tree-based ensemble method that isolates anomalies by making random selection of features, and dividing the dataset according to randomly selected thresholds. In general, anomalies tend to be isolated in fewer steps compared to normal objects, thus an anomaly score can be defined as proportional to the number of splits required to isolate a given object. IF has been used successfully for AD of astronomical time series (e.\,g., \citealt{Ishida19, Villar21, Malanchev21}).

In this work we followed both approaches to select anomalous AGN light curves: we used the VRAE model to select anomalous light curves by means of their reconstruction error, and we used the latent space as features of an IF algorithm. 

We first trained our VRAE model using the balanced phys-epochs set (the balanced phys set is used for testing purposes only). We assume that in general, the balanced phys-epochs set is composed of normal light curves, as it mostly includes type 1 QSOs. Some of the light curves in the balanced phys-epochs set could be anomalous, if they were, for instance, associated with CSAGN events, or there were problems in the ZTF PSF-fit-based photometry. However, we do not expect that these anomalous light curves correspond to a high fraction of the balanced phys-epochs set, since we are filtering the light curves using the \texttt{catflags} quality score, and since CSAGN events are very rare. 

After training, we apply the VRAE model to the full sample (230,451 sources). This sample contains not only regular type 1 QSOs, but also blazars and Seyferts, that can be used to validate the efficiency of our algorithm selecting anomalous light curves. We expect that blazars have light curves that are intrinsically different to the light curves of regular type 1 QSOs, since the optical emission is dominated by emission from the relativistic jet, and rapid time-variable beaming effects can be present \citep{Netzer13}. Seyferts on the other hand, have optical emission that correspond to a combination of the flux coming from the active nucleus, and flux coming from the host, and thus, the optical variable signal of Seyferts may look quite different to the variable signal of regular QSOs, if the contribution of the host galaxy is considerable high. They may also have systematically lower $\mathrm{BH}_{\mathrm{mass}}$ ($<10^7$) and lower accretion rate, which could lead to more rapid fluctuations (which are then damped by the host galaxy).

\begin{figure*}[tb]
\centering
\subfloat[RAE balanced phys]{
\includegraphics[width=0.48\linewidth]{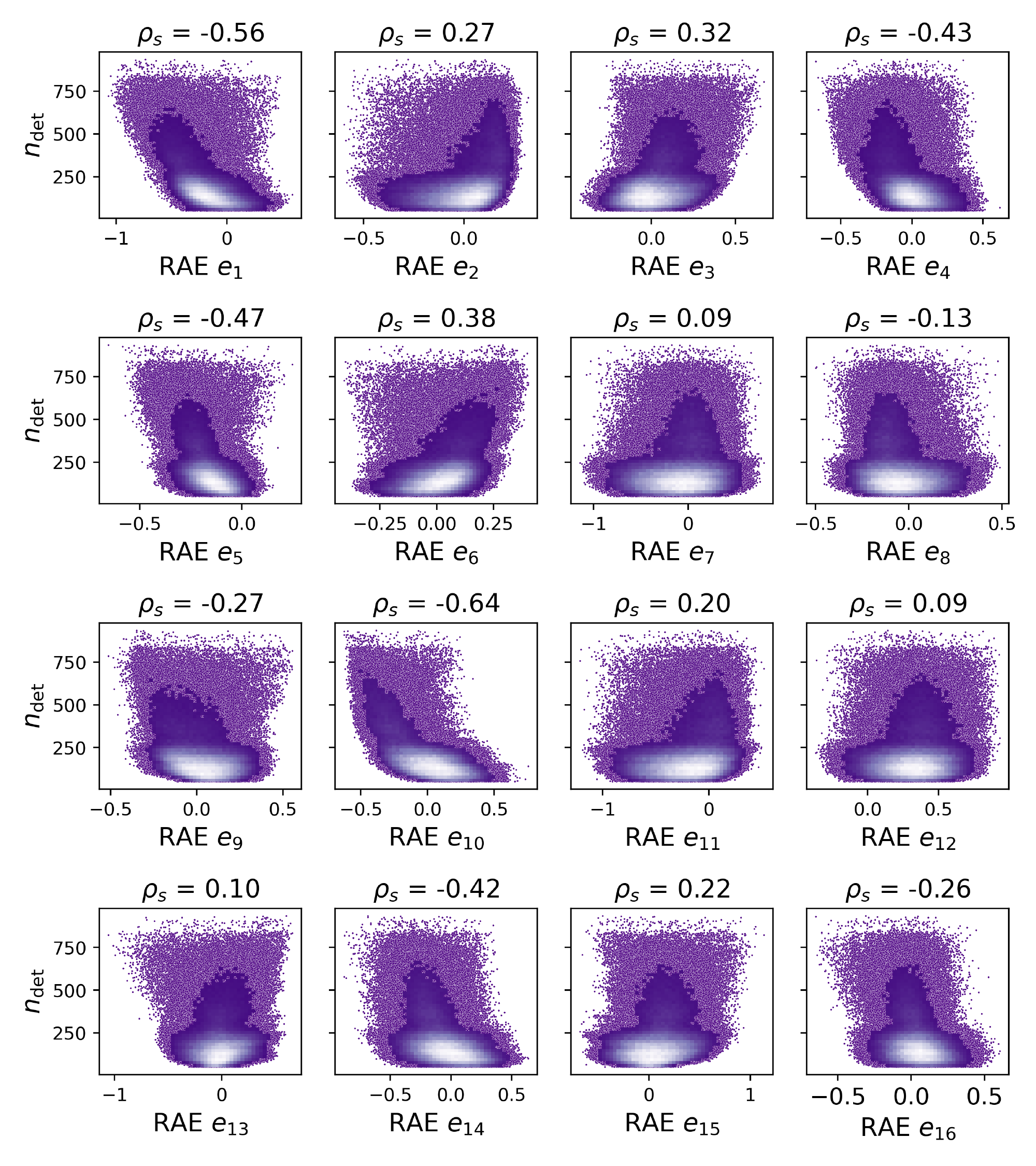}
\label{fig:space-attr-rae}}
\hfill
\subfloat[VRAE balanced phys]{
\includegraphics[width=0.48\linewidth]{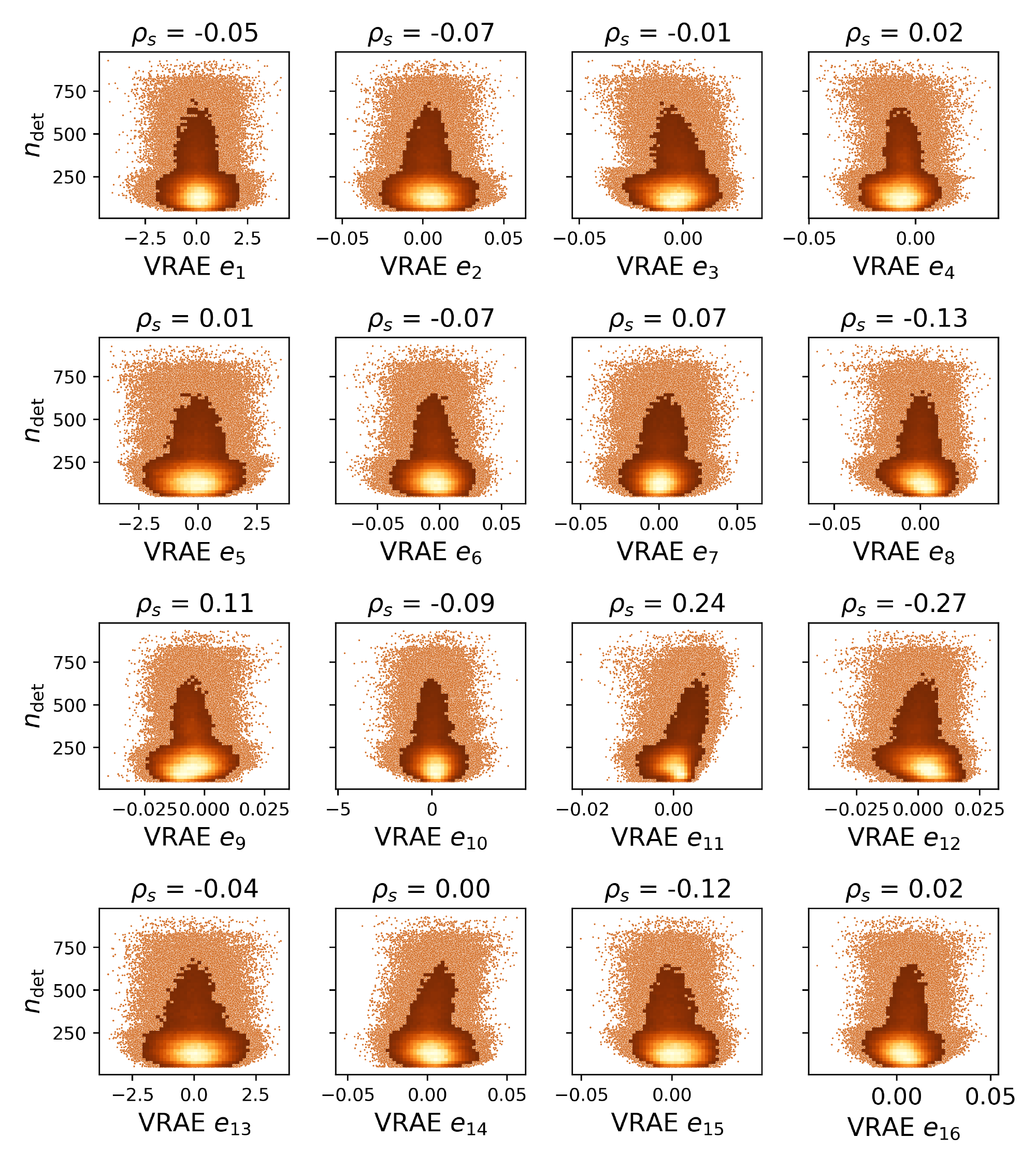}
\label{fig:space-attr-vrae}}\\
\subfloat[RAE balanced phys-epochs]{
\includegraphics[width=0.48\linewidth]{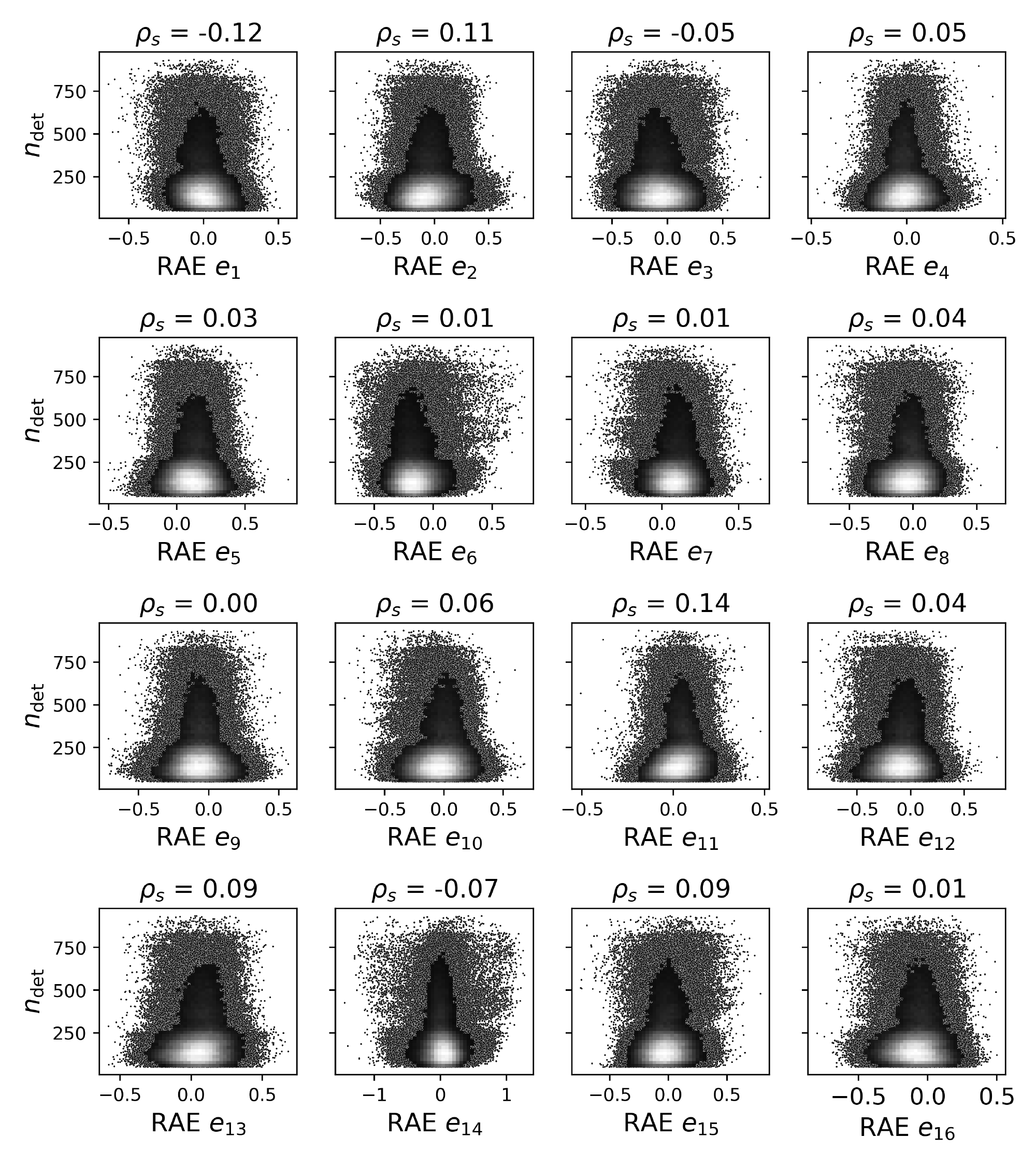}
\label{fig:AE_encoding_feats}}
\hfill
\subfloat[VRAE balanced phys-epochs]{
\includegraphics[width=0.48\linewidth]{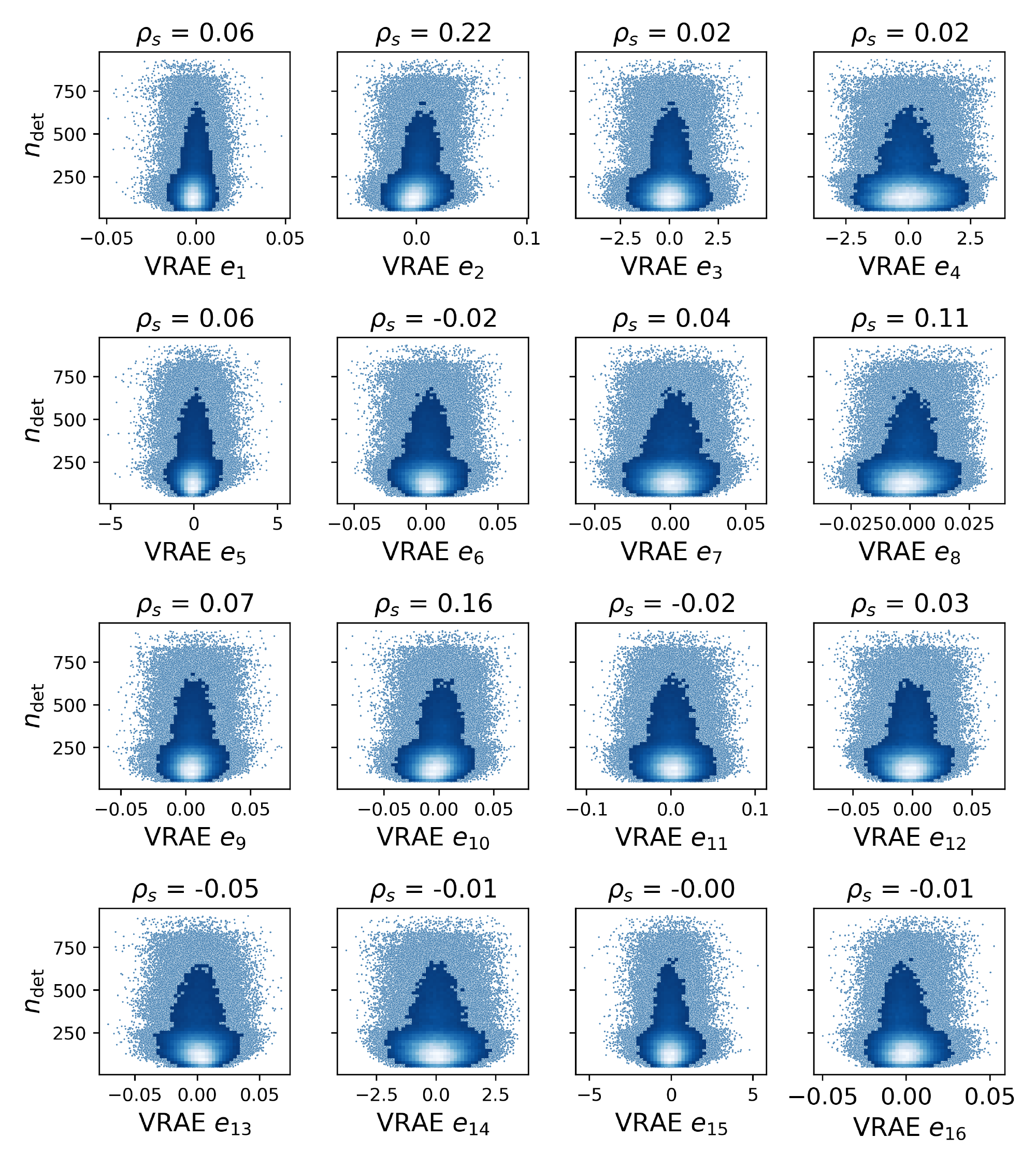}
\label{fig:VAE_encoding_feats}}

\caption{Latent space attributes as a function of the number of epochs for an RAE  (panel a; in purple) and a VRAE (panel b; in orange) architecture trained with the balanced phys set; and for an RAE  (panel c; in black) and a VRAE (panel d; in blue) architecture trained with the balanced phys-epochs set. In the four panels, brighter areas correspond to over-densities (as in Figure \ref{figure:spec_prop}, but we omitted the color bars). The Spearman rank correlation coefficient is shown as reference on top of each subplot. For all the subplots, the p-value of the coefficient is $p_{val}\leq10^{-8}$.  
\label{figure:comp_latent_space}}
\end{figure*}

To measure the reconstruction error, we sampled 10 different latent vectors, and generated 10 VRAE reconstructions for each light curve, and then computed the average reconstructed normalized light curve $\left[\,\overline{m_{rec}}\,\right]$, and the standard deviation of the reconstruction $\left[\text{std}(m_{rec})\right]$. Since the inputs of the VRAE model are the normalized light curves, we have to denormalize the VRAE reconstructions, by applying the same average magnitude and standard deviation used when we computed $m_{norm}$ and $\text{err}_{norm}$. Thus, we compute the final light curve prediction ($m_{pred}$) and its error (err$_{pred}$) as:
\begin{equation}
m_{pred}= \overline{m_{rec}}*\text{std}(m)+\overline{m}\ \; \text{,}
\end{equation}
\begin{equation}
\text{err}_{pred} = \text{std}(m_{rec})*\text{std}(m)\,.    
\end{equation}
We used as reconstruction error ($R$) the mean squared weighted deviation (MSWD; also known as the reduced chi-square statistic) 
\begin{equation}
R= \frac{1}{N_{T}} \sum_i^{N_{T}} \frac{(m_i-m_{pred_{i}})^2}{\text{err}_i^2+{\text{err}_{{pred}_i}}^2}\,.
\end{equation}

Figure \ref{figure:R_balsample} shows the reconstruction error for the sources in the balanced phys-epochs set. Fourteen percent of the sources in this sample have $R<1$, with most of them (79\%) having average ZTF photometric errors larger than 0.1, and thus their light curves are dominated by noise, which explains the low $R$ values. The peak of $R$ is around $\sim1.25$, and only a small fraction (2\%) have $R\geq3$. This is not surprising, as we can expect that a small fraction of the balanced phys-epochs set present anomalous behaviors. Thus, we used $R=3$ as our threshold to label a light curve as anomalous. Note that this value can be modify as desired to reduce or increase the number of anomalies to be presented to the final user of the method.  

In a second stage, we used the latent mean ($\mu$) as a vector of features, and selected anomalous sources with IF, using the implementation available in \texttt{scikit-learn} \citep{Pedregosa12}, with 500 base estimators, and a contamination of 2\%. The IF model was trained with the balanced phys-epochs set, and then we applied the model to the full sample. The \texttt{scikit-learn} implementation of IF provides an anomaly score (which is proportional to the mean depth of the leaf containing the observation), where the lower the value, the more abnormal is the source. We computed the IF scores ($\text{IF}_{score}$) and used the threshold value $\text{IF}_{th}=-0.57633$ (defined for a contamination of 2\% in the training set) returned by the model to label the light curves as anomalous. As for the case of $R$, this threshold value can be modified to change the number of anomalies presented to the final user, and we used here a fixed value to ease the analysis of the results obtained using our method. 

Hence, following this two-stage approach, we label a light curve as anomalous if its reconstruction error is $R\geq3$ or if its has an IF score smaller than the threshold ($\text{IF}_{score}\leq\text{IF}_{th}$). This is done because the IF scores obtained for light curves with large reconstruction errors are not trustworthy, since the latent space vectors of these light curves do not compress properly the information associated to the real variability signal. 

\section{Results}\label{sec:results}

\subsection{Comparison of the performance of the VRAE and RAE architectures}\label{subsec:performance}

\begin{figure*}[tb]
\centering
\includegraphics[width=\linewidth]{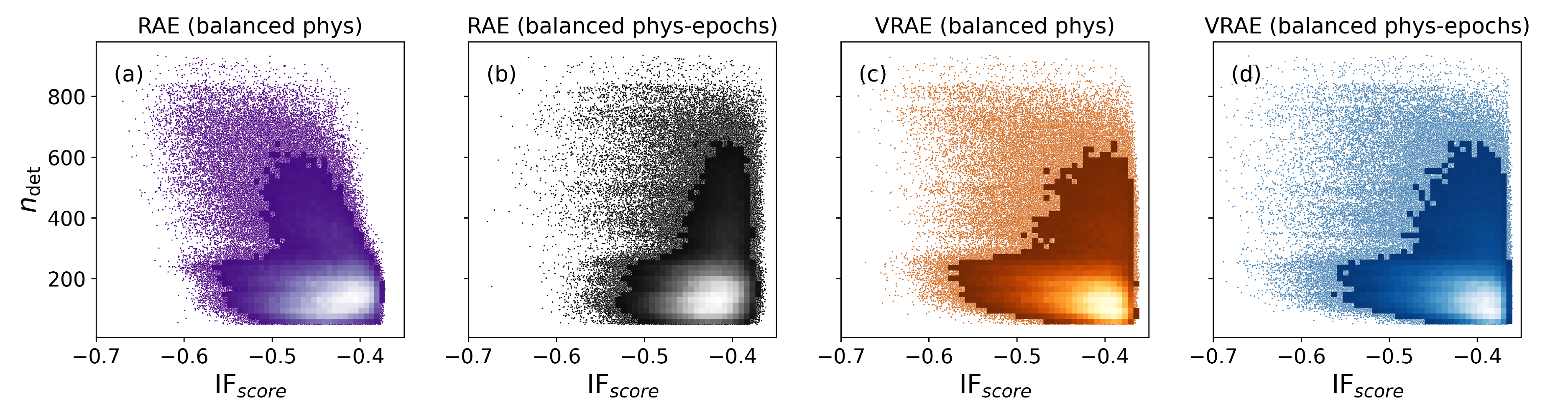}
\caption{Number of epochs per light curve as a function of the IF score for the full sample, obtained using features extracted by an RAE architecture trained with a dataset balanced by means of their physical properties (panel a; purple; RAE balanced phys), by an RAE architecture trained with a dataset balanced by means of their physical properties and number of epochs per light curve (panel b; black; RAE balanced phys-epochs), by a VRAE architecture trained with a dataset balanced by means of their physical properties (panel c; orange; VRAE balanced phys), and by a VRAE architecture trained with a dataset balanced by means of their physical properties and number of epochs per light curve (panel d; blue; VRAE balanced phys-epochs). In the four panels, brighter areas correspond to over-densities (as in Figure \ref{figure:spec_prop}, but we omitted the color bars). \label{figure:comp_IF_score}}
\end{figure*}

At the beginning of this project, we began by evaluating the anomaly detection performance of an RAE architecture similar to the one presented in \cite{Tachibana20}, using the same number of layers and units, but excluding the forecasting step. In our initial tests of that RAE architecture, we used the balanced phys set to train the model. When we applied this model to the full sample, we noticed a high correlation between the values of the attributes of the latent space (RAE $e_i$, with $i$ the number of the attribute), and the number of epochs per light curve. The panel a (purple) of Figure \ref{figure:comp_latent_space} shows the value of each attribute versus the number of epochs per light curve ($n_{\text{det}}$). The Spearman rank coefficient ($\rho_s$) is provided on top of each subplot. It is clear from the subplots and from the correlation coefficient $\rho_s$ that there is considerable correlation between some attributes and $n_{\text{det}}$, particularly for RAE $e_1$, RAE $e_4$, RAE $e_5$, RAE $e_{10}$, and RAE $e_{14}$. This effect could produce problems when we use the vector of attributes as features for an AD algorithm. When we tried to use this vector of features for the IF algorithm, most of the light curves labeled as anomalous had a large number of epochs. Panel a in Figure \ref{figure:comp_IF_score} (purple) shows the IF score obtained for this RAE architecture. It can be observed that the IF score is more negative (i.e., the sources are more anomalous) for light curves with more epochs. This happens because, as shown in Figure \ref{figure:ndet}, most of the light curves in the full sample have less than 200 epochs, and thus light curves with larger numbers of epochs are considered anomalous by the IF model.

VAEs have shown to be able to deal with AD problems properly (e.\,g., \citealt{Portillo20,Villar21}), hence we repeated the experiment using the VRAE architecture described in section \ref{subsec:vrae}, to test whether it could deal in a better way with the high imbalance in $n_{\text{det}}$ of the full sample, using the balanced phys set to train the model. The results of this test are presented in panel b of Figure~\ref{figure:comp_latent_space}, and in panel c of Figure \ref{figure:comp_IF_score} (orange plots in both figures). In this case, we used the mean latent space as a vector of attributes (VRAE $e_i$, with $i$ the number of the attribute). The values of the vector of attributes show  much less correlations in general across all the layers with $n_{\text{det}}$, with the exception of the attributes VRAE $e_{11}$ and VRAE $e_{12}$, where a weak correlation can be observed. We also used these attributes as features for our IF model (see panel c of Figure \ref{figure:comp_IF_score}), and noticed that the correlation of the IF score with $n_{\text{det}}$ is highly reduced.  These results tell us that the proposed VRAE architecture is better than the original RAE architecture not only because of its smooth latent space representation, but also because it offers a latent space robust against diversity in the length of the time series; basically more motivated by physical differences than by those in the observing strategy.

Finally, we tested whether the correlation with the $n_{\text{det}}$ is eliminated when we train both RAE and VRAE architectures with a training and validation sets that are balanced by means of their physical properties and $n_{\text{det}}$. We used the balanced phys-epochs set described in Section \ref{subsec:traningsets}. The results for the full sample when using the RAE architecture are shown in panel c of Figure \ref{figure:comp_latent_space}  and in panel b of Figure \ref{figure:comp_IF_score} (black plots in both figures). We can see in both figures that the correlation with $n_{\text{det}}$ is highly reduced. A similar result is observed for the VRAE architecture, whose results are shown in panel d of Figure~\ref{figure:comp_latent_space} and in panel d of Figure~\ref{figure:comp_IF_score} (blue plots in both figures). In general, the correlation with $n_{\text{det}}$ is nonexistent or very weak (e.\,g., VRAE $e_2$, and VRAE $e_{10}$), although the improvement with respect to the previous VRAE model is small, and the main differences are observed in the IF$_{score}$ obtained for light curves with $n_{\text{det}}>600$ (see panels c and d of Figure \ref{figure:comp_IF_score}). 

From these results, we can conclude that a VRAE architecture provides better performance when dealing with datasets that are highly imbalanced in their observational properties, like $n_{\text{det}}$. And that for both RAE and VRAE architectures better results can be obtained when balancing the training and validation sets by means of $n_{\text{det}}$. We decided to use the VRAE architecture trained with the balanced phys-epochs set as our final model, since it provides results that are not biased by the number of epochs, and because the smooth latent space facilitates the detection of anomalies. 

\subsection{Quality of the reconstructed light curves}\label{subsec:reconstructions}

\begin{figure}[tb]
\centering
\includegraphics[width=\linewidth]{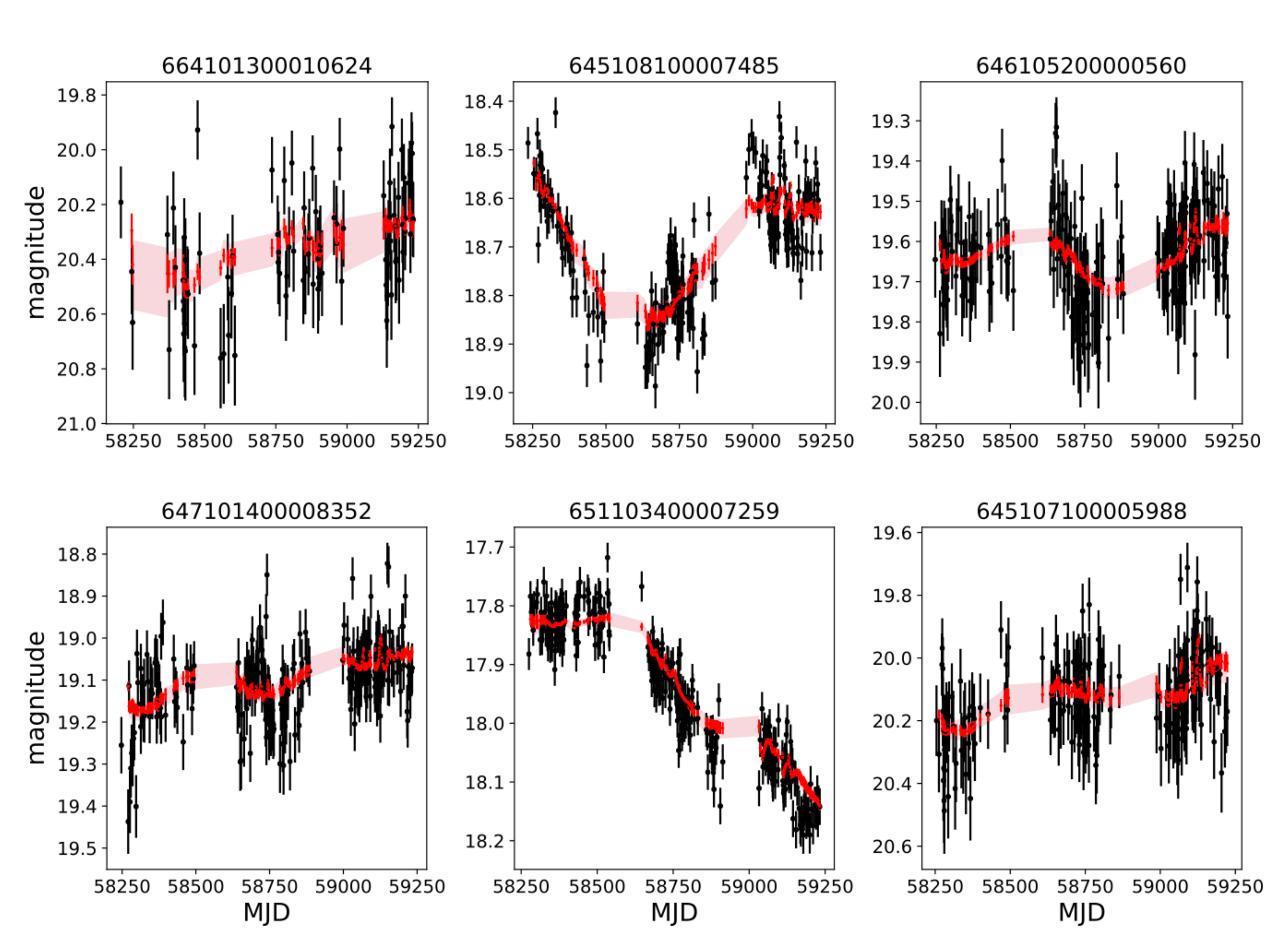}
\caption{Original light curves (black) and reconstructed light curves (red) for a random selection of light curves from the full sample.}
\label{figure:rand_lc_samp}
\end{figure}

\begin{figure}[tb]
\centering
\includegraphics[width=\linewidth]{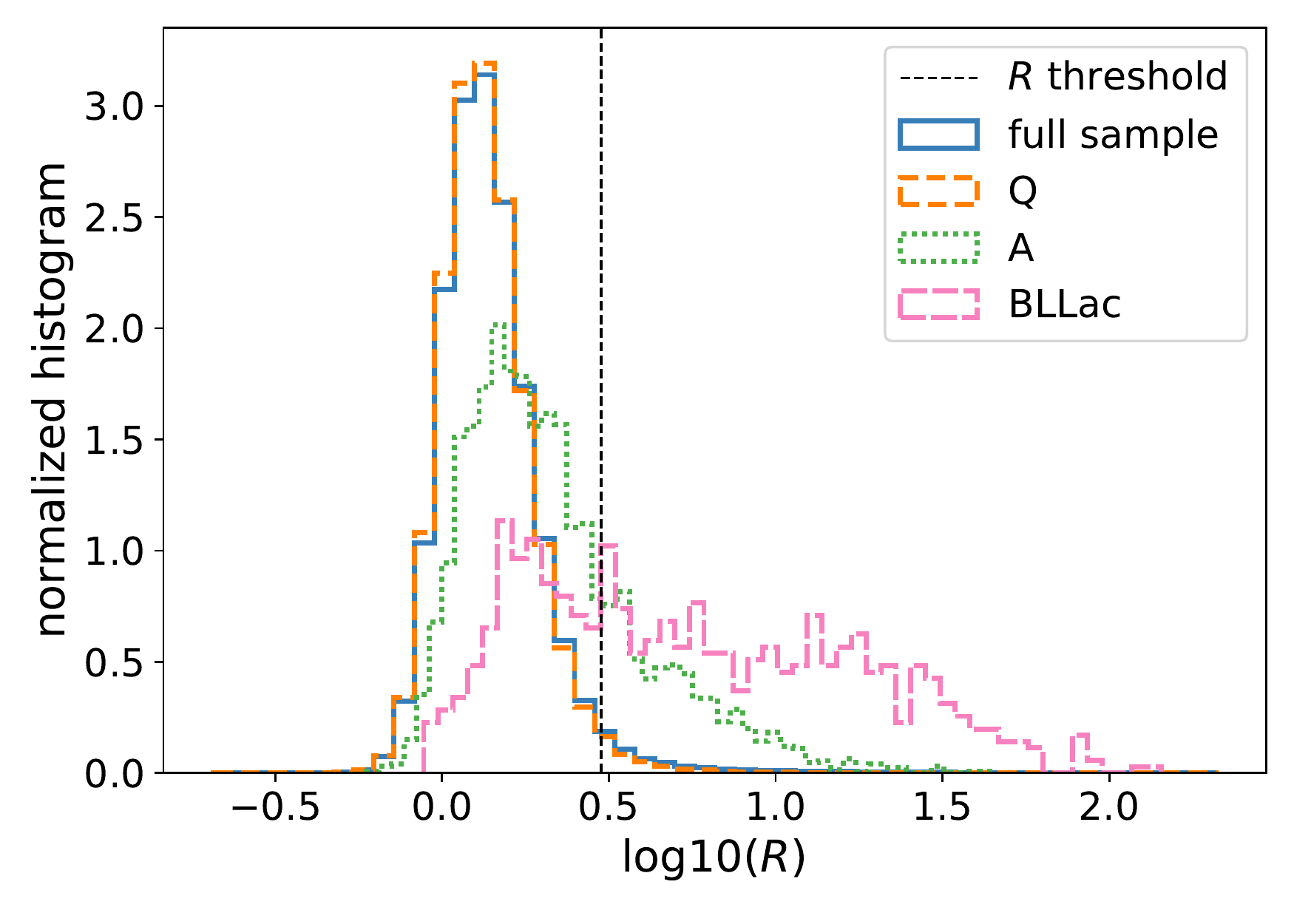}
\caption{Normalized histogram of the reconstruction error ($R$) per light curve for the full sample (blue solid), the Q class (orange dashed), the A class (green dotted), and the BLLac class (pink densely dashed). The threshold of $R=3$ is shown with a black dashed line. \label{figure:R_fullsample}}
\end{figure}

We used the VRAE architecture trained with the balanced phys-epochs set to reconstruct all the light curves in the full sample. Figure \ref{figure:rand_lc_samp} shows a random selection of original light curves and reconstructions from the full sample. Figure~\ref{figure:R_fullsample} shows the normalized histogram of the reconstruction errors obtained for the light curves in the full sample, and  light curves with classes Q, A, and BLLac. Most of the sources in the full sample have class Q (see Figure~\ref{figure:num_classes_ls}), thus is not surprising that the distribution of $R$ for the full sample is quite similar to the distribution of the class Q. 

Thirteen percent of the total sample have $R<0.1$. Similar to the balanced sample, a high fraction of these sources have ZTF light curves with large photometric errors (82\% have average photometric errors larger than 0.1). After a visual inspection, we noticed that several of them seem to have overestimated photometric errors, which explain the low $R$ values measured for them.

The high $R$ values obtained for BLLacs are not surprising, as these sources are expected to be intrinsically different to regular quasars. The high $R$ values obtained for some objects with the class A could be produced by three factors: 1) contamination from the host is damping the light curves; 2) there are still some issues with the photometry, even though we removed from the sample the more extended sources (e.g., problems in the photometry not identified by the ZTF pipeline); and 3) the light curves of class A and Q sources are intrinsically different, due to e.g., different physical properties ($\mathrm{BH}_{\mathrm{mass}}$, accretion rate, redshift). The observed results are probably due to a combination of these three factors, since the lower luminosity leads to more host galaxy contribution, and also host-dominated sources normally appear as extended sources in the ZTF images.

By visually inspecting the reconstructed light curves we noticed that long-term variations are properly recovered by the reconstructions, however, in some cases, more rapid variations are missed by the reconstruction. An example of this can be observed in the top middle panel of Figure \ref{figure:rand_lc_samp}. \cite{Tachibana20} observe a similar result for the RAE architecture. This may be related to the cadence and photometric noise of the light curves, but it could also reflect the fact that in general sources belonging to class Q do not show very rapid variations (the timescale of the variations is of the order of months or years, since they are more luminous and more massive). The results obtained for class A sources are probably explained by this, since they are expected to show more rapid variations, as they are less luminous and less massive \citep{MacLeod10,Simm16,Sanchez-Saez18}. 

In any case, the results obtained by the VRAE architecture are quite robust, regardless of the diversity of cadences, photometric errors, and variability behaviors, and the fact that we are neither interpolating the data, nor correcting by redshift, nor doing any light curve manipulation, as other works that use a similar architecture do (e.\,g., \citealt{Naul18,Villar20,Villar21}).

\subsection{Selection of anomalous light curves}\label{subsec:an_lcs}

As described in Section \ref{subsec:ad}, we applied our trained VRAE model to the full sample, and used the mean of the latent space as attributes for our IF algorithm. Then we classified a given light curve as anomalous if its VRAE reconstruction error is $R\geq3$ or if its $\text{IF}_{score}\leq\text{IF}_{th}$.

\begin{figure}[tb]
\centering
\includegraphics[width=\linewidth]{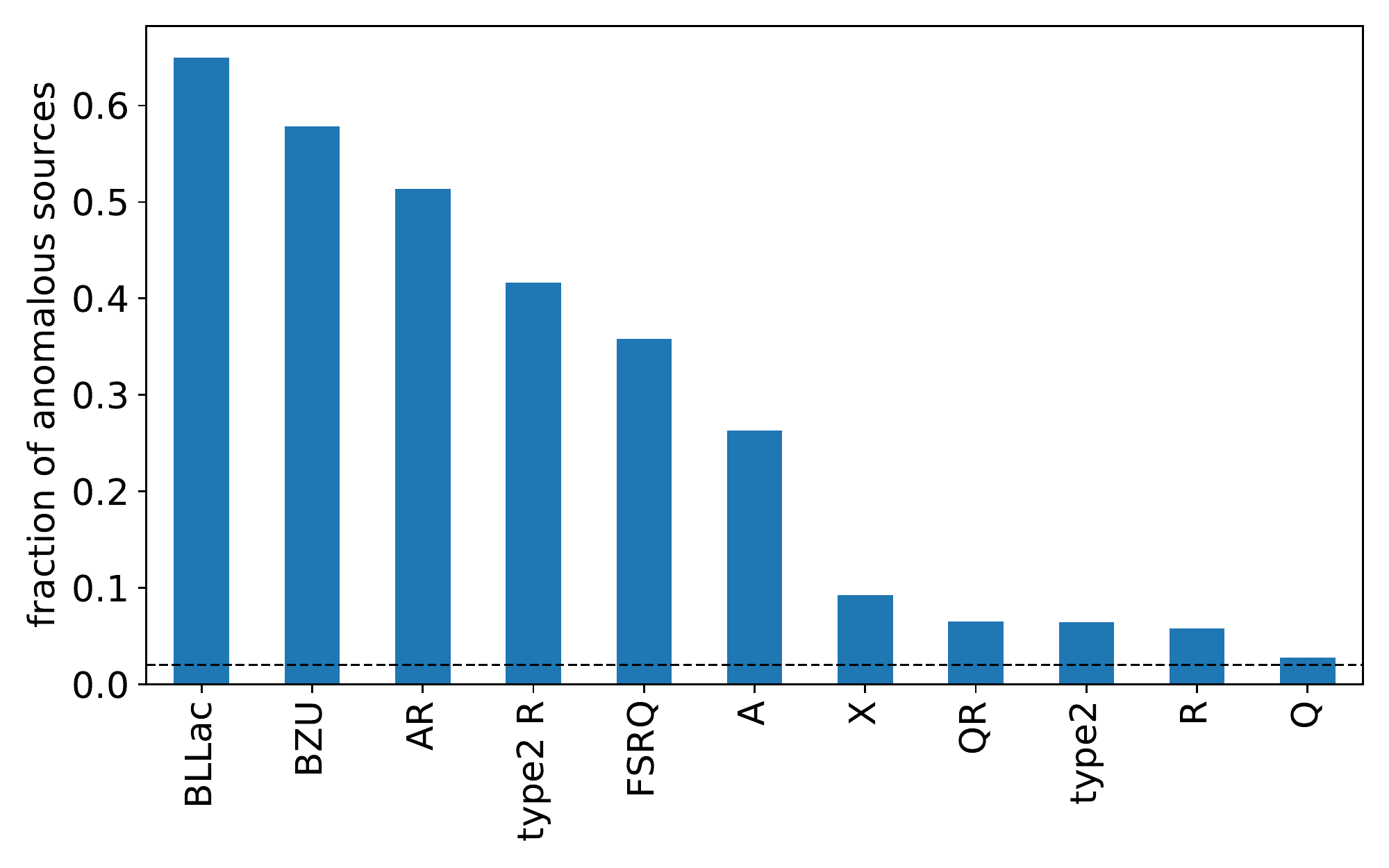}
\caption{Fraction of light curves per class classified as anomalous. The black dashed line highlights the position where the fraction is equal to 2\%.}
\label{figure:anomalous_fraction}
\end{figure}

Figure \ref{figure:anomalous_fraction} shows the fraction of light curves labeled as anomalous for each of the classes described in Section \ref{subsec:catalogs}. The low fraction of anomalous objects belonging to the class Q is not surprising, since we trained our model with a dataset that is dominated by this class. The fraction of anomalous Q sources is close to the 2\% that we defined as threshold in our AD strategy. On the other hand, the classes with the largest fraction of anomalies are the three blazar classes (BLLac, BZU, and FSRQ), the host-dominated classes (A and AR), and the type 2 with radio emission (type2 R). 

As mentioned in Section \ref{subsec:catalogs}, blazars were included in the analysis to test the efficiency of our AD methodology, and we expected these classes to be classified as anomalous. The results obtained for blazars therefore confirm the capacity of our model to detect anomalous AGN light curves. Interestingly, the fraction of anomalous FSRQ is $\sim35\%$, and thus much smaller than the fraction observed for BLLacs ($\sim65\%$). From this we might conclude that FSRQs have optical emission that is intrinsically more similar to the optical emission of regular quasars, compared to BLLacs, but further analysis is required to confirm this result. 

Something interesting that can be noticed from Figure \ref{figure:anomalous_fraction} is that in general the classes with detected radio emission (AR, type2 R, and QR), tend to have a larger fraction of light curves classified as anomalous, compared to their counterpart classes without detected radio emission (A, type 2, and Q). From this we might conclude that AGNs with radio emission have intrinsically different optical variability behaviors compared to AGNs without radio emission. Further analysis is required to confirm this result, as we are using the classifications provided by MILLIQUAS, and thus we cannot ensure that all the sources without radio emission do not emit in the radio wavebands.

\subsection{Types of anomalies found}\label{subsec:anomalies}

In this work we used the AD methodology as a recommendation system to search more easily for CSAGN candidates. Thus, in order to select CSAGN candidates, we visually inspected the sources classified as anomalies by our method, excluding the three blazar classes, and the class R, since these classes were included only for testing purposes. We define a source as a promising CSAGN candidate if its light curve shows during the visual inspection evidence of flaring activity, or abrupt increase/decrease of the optical flux. By doing this selection, we cannot ensure that the candidates are in fact CSAGNs, as they can also correspond to examples of atypical flaring activity (e.\,g., \citealt{Graham17,Trakhtenbrot19NatAs,Frederick20}), or extremely variable AGNs (e.\,g., \citealt{Rumbaugh18,Guo20,Luo20}), and that is why we called them promising CSAGN candidates.

Here we provide a general description of the types of anomalies found for the classes Q, QR, A, AR, type2, type2 R, and X. A sample of promising CSAGN candidates is presented in Section~\ref{subsec:candidates}.

\begin{figure*}[tb]
\centering
\includegraphics[width=\linewidth]{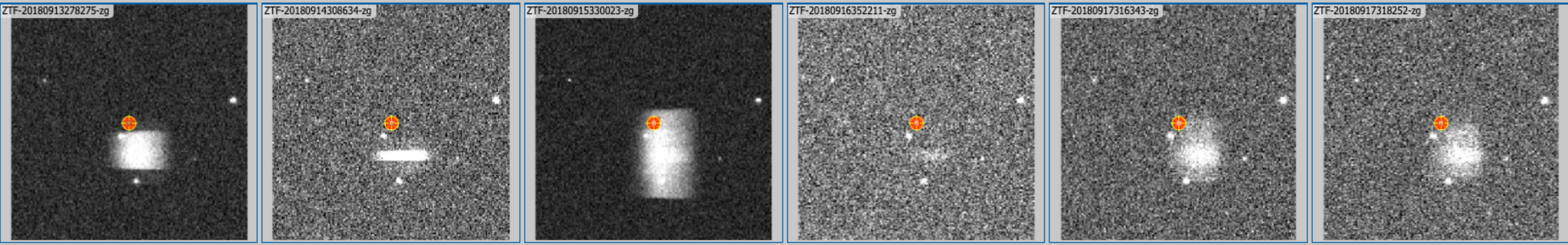}
\includegraphics[width=\linewidth]{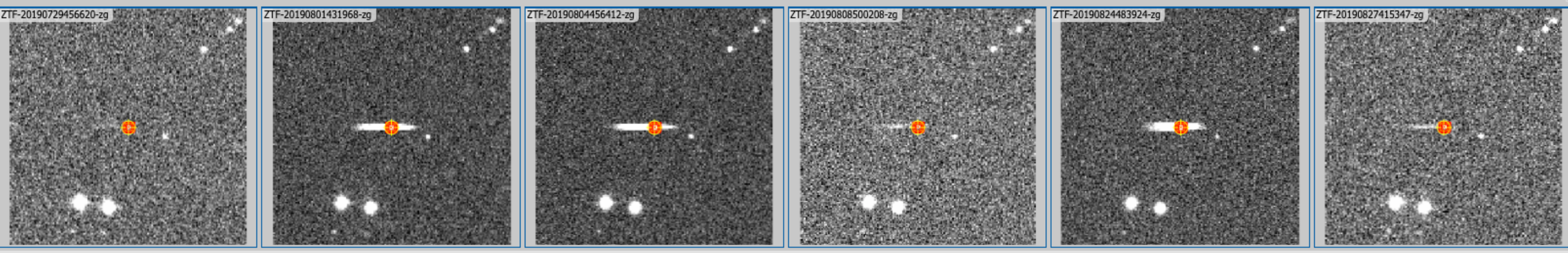} 
\caption{Set of ZTF stamps (with a cutout size of 3 arcminutes) in the $g$ band for two anomalies with class Q, with ZTF DR5 ID 498110300023393 (top) and 646103300003619 (bottom). The orange circles show the location of the target in the stamps. The stamps were obtained using the ZTF Time Series Tool. 
\label{figure:bad_stamps}}
\end{figure*}

\begin{itemize}
    \item \textit{Class Q (5774 anomalies):} around 40\% of the anomalies have very noisy light curves, where the reported photometric errors seem to be underestimated (since they show outlier observations with small photometric errors), or show evidence of problems in the photometry. We inspected the ZTF stamps (using the ZTF Time Series Tool) of some of these sources and found that in some cases there are artifacts in the images, that are not removed when using \texttt{catflags}$=0$. Figure \ref{figure:bad_stamps} shows a set of stamps for two of these candidates. All these stamps have \texttt{catflags}$=0$, however the artifacts in the images are easily identified. 
    In addition, the anomalies in this class have a peak in redshift at $z\sim0.5$, while the Q sources labeled as normal have in general $z>1$, and thus, is probable that a high fraction of the anomalies have light curves distorted by the emission of their host galaxies, or are affected by time dilation and rest frame differences. Moreover, we found two sources that are in fact variable stars, but were classified as quasars by the SDSS pipeline (ZTF DR5 ID 794110400002914 and 801111300003144). We identified 49 promising CSAGN candidates.
    \item \textit{Class QR (747 anomalies):} a high fraction of the anomalies have variability behaviors similar to those observed for BLLacs. In particular, 63\% of the QR anomalies have classifications in the ALeRCE light curve classifier \citep{Sanchez-Saez21}, and 26\% of these are classified as blazars. In addition, 36 of these anomalies are classified as BLLac in the SIMBAD database \citep{Wenger00}. After the visual inspection we identified 10 promising CSAGN candidates.
    \item \textit{Class A (873 anomalies):} the anomalies in this class have lower redshifts than those labeled as normal, and therefore are probably highly dominated by the emission from the host galaxies. Their light curves show in general low-amplitude rapid variations, which are not properly recovered by the VRAE reconstructions. We found one source classified as AGN by the SDSS pipeline (ZTF DR5 ID 408114200003141), but classified as YSO in SIMBAD. We identified 10 promising CSAGN candidates.
    \item \textit{Class AR (190 anomalies):} as with class A, the anomalies in this class have lower redshifts than those labeled as normal, and thus, their light curves could be more contaminated by the emission from the hosts. Similar to the QR class, some sources show variability behavior close to those observed for BLLacs. 88\% of these anomalies have classification in the ALeRCE light curve classifier, and 18 are classified as blazar. In addition, 10 are classified as blazar in the SIMBAD database. We identified three promising CSAGN candidates.  
    \item \textit{Classes type2 and type2 R (20 anomalies):} several of the type2 anomalies are probably type 1.8 or 1.9 sources, since their SDSS spectra show evidence of a broad emission line component, which may help to explain their optical variations. We identify two promising CSAGN candidates. 
    \item \textit{Class X (246 anomalies):} when we inspected the light curves of these anomalies, we noticed that several of them seem to be variable stars. 46\% of the anomalies have classification in the ALeRCE light curve classifier, and 87 are classified as variable star (including several cataclysmic variables and young stellar objects). In addition, 65 have classification in the Simbad database, and only four of them are classified as AGN or AGN candidate, the rest are labeled as variable stars. We identify one promising CSAGN candidate.
\end{itemize}

\subsection{CSAGN candidates}\label{subsec:candidates}

To select the sample of CSAGN candidates we rejected very noisy light curves, the sources identified as variable stars in SIMBAD, and the sources that can be classified as variable stars according to their light curves. We selected as promising CSAGN candidates those anomalies that present evidence of flares, and/or abrupt  increment  or  decrement  in  the  luminosity.

We identified 75 promising CSAGN candidates (65\% of them belong to the class Q), 73 of these have alerts in the ZTF alert stream. The full list of candidates is presented in Appendix \ref{appendix:cand_list}. Four of these candidates appear in the list of CSAGNs presented in \cite{Graham20}, and another two have been confirmed spectroscopically (Graham, private communication). Moreover, 28 of the candidates have been selected as CSAGN candidates using CRTS data and/or ZTF DR3 data (Graham et al. in prep.). All these candidates are highlighted in Appendix \ref{appendix:cand_list}.

Sixty candidates have at least one SDSS spectrum available, with 12 observed relatively recently (MJD$>58000$) such that it overlaps with the overall baseline of the ZTF light curves. We leave for future work any detailed spectroscopic analysis of the sample, as this type of study is beyond the scope of this work, and we provide here a brief description of the properties of the most interesting sources. 

\begin{figure*}[tb]
\centering
\includegraphics[width=\linewidth]{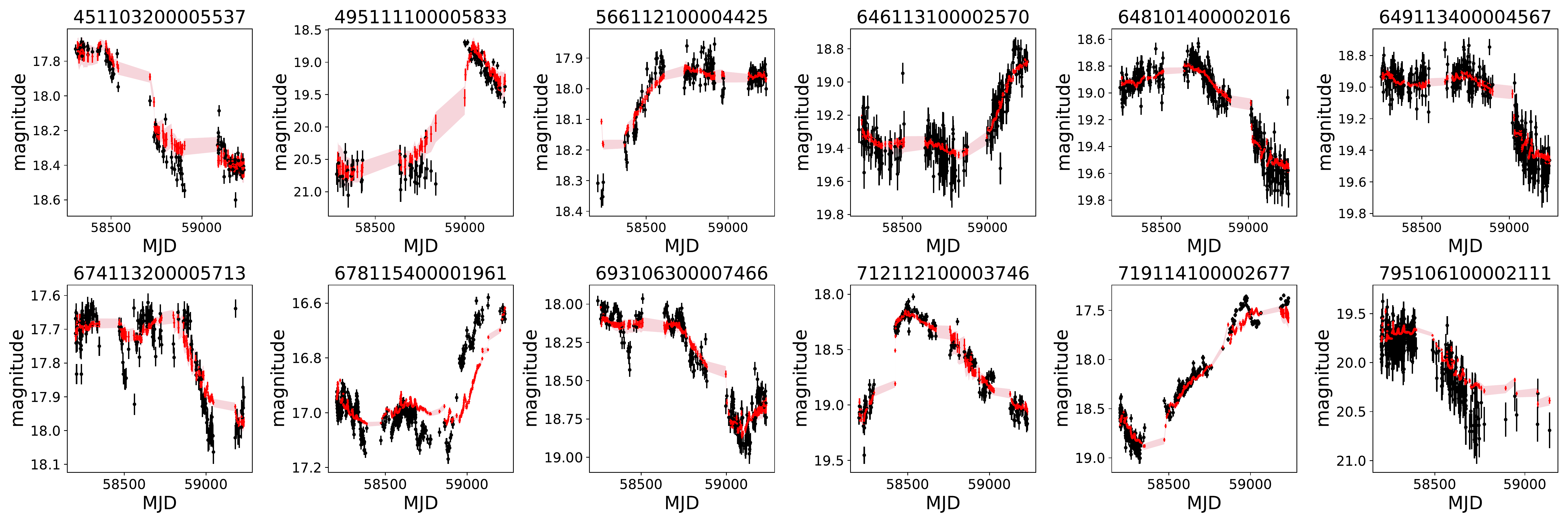}
\caption{As in Figure \ref{figure:rand_lc_samp}, but for CSAGN candidates with recent SDSS spectra. For the reconstruction we show the average and standard deviation (error bars) of the 10 reconstructions. The shaded regions show the minimum and maximum values obtained for the reconstruction at each epoch. \label{figure:cand_lc_examp}}
\end{figure*}

The light curves and reconstructions of the 12 candidates with recent SDSS spectra are shown in Figure \ref{figure:cand_lc_examp}. We can see that the light curves show a variety of shapes. We noticed that 72\% of the total sample of candidates present periods of low amplitude variations (sometimes even constant emission), preceded or followed by abrupt variations, and we called this the ``plateau'' feature (e.\,g., ZTF DR5 ID 566112100004425, 646113100002570, and 649113400004567). These type of light curve shapes have been previously detected in X-ray light curves of X-ray binaries, and are associated to changes in the accretion state (e.\,g., \citealt{Remillard06,Done07,Sobolewska11}), and also recently in X-ray light curves of CSAGNs during outburst activity \citep{Ruan19b}. 

Five of the sources with recent SDSS spectra show very narrow H$\alpha$ profiles (ZTF DR5 IDs: 674113200005713, 678115400001961, 693106300007466, 712112100003746, and 719114100002677). One of these (ZTF DR5 ID 674113200005713) was previously classified as a Narrow-line Seyfert 1 (NLS1) by \cite{Rakshit17nls1}. Additionally, three of the candidates with old SDSS spectra (MJD$<58000$) are classified as NLS1 in \cite{Rakshit17nls1} (ZTF DR5 IDs: 479108100002617, 582108200000456, and 676103300001128). Previous works have shown that NLSy1 can present rapid enhanced flaring activity (e.\,g., \citealt{Miller00,Frederick20}), and thus we need to follow-up the candidates to confirm whether they correspond to CSAGN events, or are examples of these new optical transients detected in NLSy1. 

Finally, there are nine candidates with atypical spectra (ZTF DR5 IDs: 427114100001870, 474105300001176, 526115100009333, 578114100000128, 616105200002737, 626104200000260, 664111300002498, 789103200005093, and 818114300000932), with two of them (ZTF DR5 IDs: 626104200000260, and 664111300002498) showing very asymmetric broad line profiles, similar to those observed in NGC 3516, a known CSAGN \citep{Oknyansky21}.

Due to the current COVID-19 pandemic and subsequent lengthy closure of the observatories, we have not been able to follow-up our candidates, but we provide to the astronomical community the candidate's list so that future follow-up campaigns can target them.

\section{Discussion} \label{sec:discussion}

\subsection{About the use of AD techniques to search for CSAGNs}\label{subsec:use_AD}

In this work we used an AD approach to search for CSAGNs.  A high fraction of the anomalies found by our method are bogus candidates (e.\,g., noisy light curves, or light curves with problems in the photometry), or sources with incorrect labels in the original catalogs (e.\,g., variable stars classified as quasars by the SDSS pipeline, or X-ray detected variable stars). This type of issue has been found in previous works that used AD algorithms to search for anomalous light curves (e.\,g., \citealt{Malanchev20}, P\'erez-Carrasco et al. in prep). Thus, our algorithm can be a powerful tool to identify light curves with photometric issues in massive datasets, and to identify mislabeled sources.

Moreover, as mentioned in Section \ref{subsec:reconstructions}, the light curves of host-dominated AGNs are not properly reconstructed by our VRAE architecture, producing a large fraction of anomalous candidates, which hinders the discovery of true outliers such as CSAGNs from this population. \cite{Tachibana20} found similar results and decided to remove host-dominated AGNs from their analysis. In our case, we were able to detect promising CSAGN candidates in the A and AR classes, and thus, for the sake of inclusivity, we do not rule out host-dominated AGNs from this work. A possible solution to this problem would be to remove the host galaxy emission from the light curves, or to use light curves obtained from difference images. We leave this type of analysis for future work.   

These results show us that we cannot blindly use the output of our AD algorithm to define a sample of CSAGN candidates, and demonstrate that interaction between AD algorithms with domain experts is needed in order to find reliable samples of candidates (\citealt{Ishida19,Malanchev20}; P\'erez-Carrasco et al. in prep). 

In addition, the candidates presented in this work require further confirmation, as atypical variations have also been observed in other AGN processes, like flaring activity in NLS1s \citep{Trakhtenbrot19NatAs,Frederick20}, or extremely variable  AGNs \citep{Rumbaugh18,Guo20,Luo20}. 

Finally, it is encouraging that we are finding promising CSAGN candidates in light curves with timespans that range between 2 and 2.8 years. We expect that much better results will be obtained by our AD algorithm in the future with the upcoming ZTF data releases, and also with data obtained by the Vera Rubin Observatory Legacy Survey of Space and Time (LSST; \citealt{LSST}), or by combining different data sets, to extend the baselines by $\sim1$ order of magnitude and in different filters. 

\subsection{The plateau feature}\label{subsec:plateau}

A high fraction (72\%) of our CSAGN candidates show light curve shapes with a fast rise or decay and a ``plateau''  that spans hundreds of days, and we called this shape the ``plateau feature'' (e.\,g., sources with ZTF DR5 ID 566112100004425, 646113100002570, and64911340000456 in Figure \ref{figure:cand_lc_examp}). These light curve shapes are reminiscent of those previously observed in X-ray binaries at X-ray energies \citep{Remillard06,Done07,Sobolewska11}. In these type of sources, it is well established that there exists a coupling between the accretion disc and the jet, giving rise to different states that can be followed up on human timescales thanks to the small size of the black holes \citep{Fender04}. It has been suggested that AGN might be scaled-up versions of X-ray binaries, and thus the different states could also be present \citep{McHardy06}, but with subsequently longer timescales between them.

If we extrapolate the accretion behaviors observed in X-ray binaries to regular AGNs \citep{Fender04}, the predicted timescales of these accretion state transitions are ${\sim}10^4$--$10^5$ years (if the changes are related to viscous timescales), and thus much larger than the ones observed in our CSAGN candidates. However, these state transitions have been observed in CSAGN events in the X-ray domain \citep{Ruan19b,Ricci20}, and thus it is possible to use CSAGNs to obtain insights about the similarities between the accretion flows of X-ray binaries and AGNs (e.\,g., \citealt{Ruan19}). It remains an active area of investigation as to what is the physical mechanism behind this plateau feature, and we leave that for future analysis.

\subsection{Comparison with previous works}\label{subsec:previous_works}

\begin{figure}[tb]
\centering
\includegraphics[width=\linewidth]{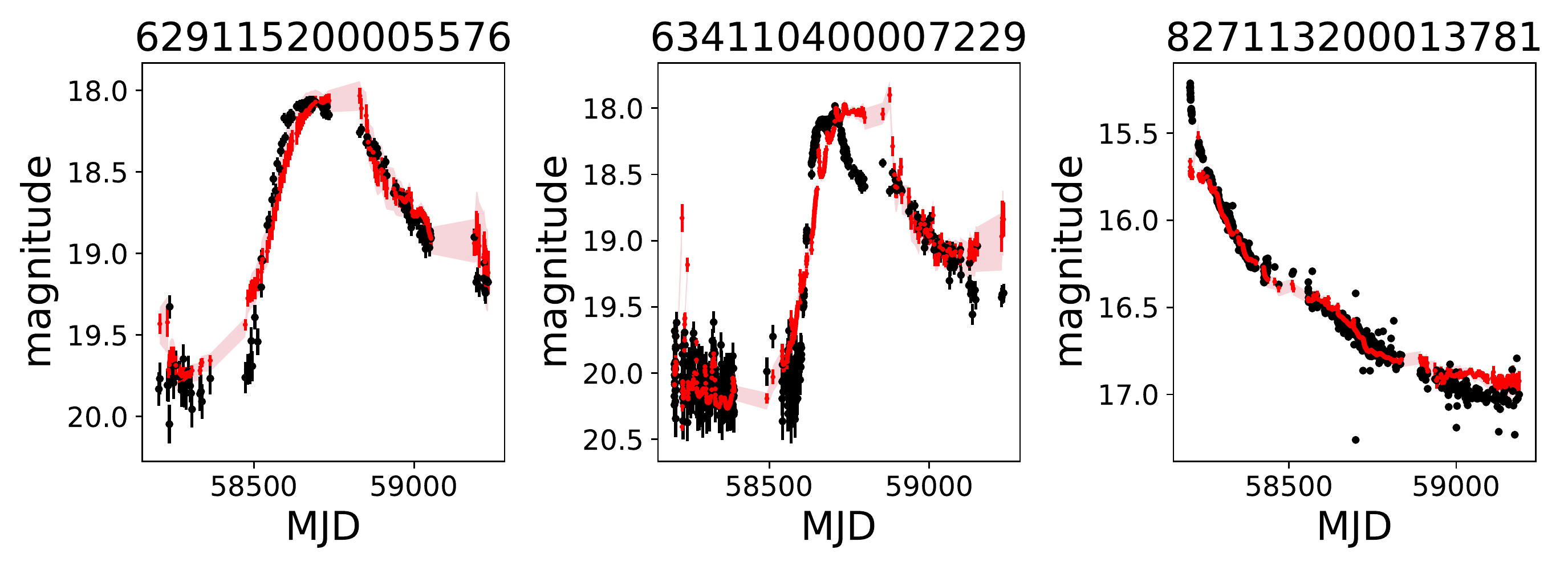}
\caption{As in Figure \ref{figure:rand_lc_samp}, but for two flaring events presented in \cite{Frederick20} (ZTF DR5 IDs 629115200005576 and 634110400007229), and for the CSAGN event presented in \cite{Ricci20} (ZTF DR5 ID 827113200013781). \label{figure:others_cand}}
\end{figure}

\cite{Frederick19} and \cite{Ricci20} detected CSAGN events (six and one event, respectively) that are visible in the ZTF DR5 data, and \cite{Frederick20} presented a sample of five NLSy1 detected in the ZTF alert stream, with atypical flaring activity. None of these sources were included in our analysis, since all of them are classified as extended sources by \cite{Tachibana18} (\texttt{ps\_score}$<0.5$). Despite this, we applied our AD methodology to these sources, and all of them were classified as anomalies. The ZTF PSF-fit-based light curves of these targets are quite noisy, and thus we consider their detection with our methodology with caution, except for three sources: the CSAGN event presented in \cite{Ricci20}, and two flaring events from \cite{Frederick20}, whose light curves are shown in Figure \ref{figure:others_cand}. These sources have very well sampled light curves in the ZTF alert stream, and variability amplitudes larger than 1 magnitude, and thus, were easy to identify using only the alert stream. Our method on the other hand, allows the detection of anomalous variations with any level of variability amplitudes, which favors the early detection of CSAGN events, and the detection of CSAGN events with smaller amplitudes, that could be missed by the ZTF alert stream. 

\cite{Graham17,Graham20} used more than eight years of CRTS data to search for AGNs with extreme optical variability, finding a sample of 51 sources with major flares, and 111 CSAGN candidates, respectively. \cite{Graham17} used a Weibull distribution to characterize the flares, and \cite{Graham20} used BB representation \citep{Scargle13} to identify significant changes of variability behaviors. The large baseline of the CRTS data allowed them to use these techniques to find extreme variability behaviors with timescales of several years, which are still impossible to detect with the existent baseline of the ZTF data, and thus, we do not consider applying such techniques to the current ZTF light curves. With the upcoming ZTF data releases, we might be able to complement our AD methodology with these other techniques. 

\cite{MacLeod16} used SDSS and Pan-STARRS1 (PS1; \citealt{Chambers16}) data to select CSAGN candidates. They used a sample of quasars with repeated SDSS spectra, and defined as CSAGN candidates those sources with large photometric variations ($|\Delta g|>1$). The use of a sample of well characterized quasars avoided the introduction of contamination from incorrect classifications. However, this way to select CSAGN candidates based solely on their variability amplitudes cannot be extrapolated to larger datasets, since the contamination from noisy light curves, or from sources with incorrect labels (e.\,g., variable stars, or even blazars), would be considerably larger. 

\section{Conclusions} \label{sec:summary}

In this work we presented an AD algorithm to search for AGN with anomalous light curves in massive datasets (like ZTF DR5), whose main aim is to find CSAGN candidates. We used an architecture inspired by the work with RAE architecture presented by \cite{Tachibana20}, but with two modifications: we used GRUs instead of LSTMs, and we used a VRAE architecture, instead of RAE. 

Our dataset is composed of 230,451 $g$ band light curves from ZTF DR5. The light curves are associated with different classes of AGNs, although 90.7\% are regular quasars (class Q). The rest of the sources are type 2 AGNs, Seyfert galaxies, blazars, and sources selected by their X-ray or radio emission. We used the ZTF PSF-fit-based light curves from DR5, with more than 2 years of data, and more than 50 epochs.

We first tested the original RAE architecture and noticed that the attributes of the latent space were highly correlated with the number of epochs per light curve, particularly when the training and validation sets were not balanced by means of the number of epochs (using the balanced phys set). We therefore tested whether a VRAE architecture is able to solve this issue, finding a considerable reduction in the level of correlation with the number of epochs, when using the latent mean of the VRAE as a vector of attributes. We decided to train the RAE and VRAE architectures with training and validation sets that were balanced by means of their spectroscopic properties (obtained from SDSS data), and by means of the number of epochs per light curve (using the balanced phys-epochs set), to avoid any correlation with the cadence of the light curves. We found that when using the balanced phys-epochs set, the latent space of the RAE architecture reduced the correlations with the number of epochs considerably, while for the case of the VRAE architecture there is a small improvement. From these results, we decided to use the VRAE architecture trained with the balanced phys-epochs set for our AD algorithm, as a VRAE architecture have the potential of solving any other balancing issues present in the data (e.g., redshift, host galaxy contamination, presence of a jet, among others). This allowed us to work with the varied cadences present in the ZTF light curves, which have large differences in the number of epochs per light curve, and also allowed us to use the latent space as a vector of features for an IF algorithm. 

We used a two-stage approach to classify anomalies. First we reconstructed all the light curves in our sample with our VRAE architecture, and measured the reconstruction error $R$ for each light curve. We classified as anomalies all the light curves with $R\geq3$. Then, we used the mean latent space as features for an IF algorithm, and classify as anomalies all the sources with an IF score smaller than the threshold defined for a contamination of 2\% for the training set ($\text{IF}_{score}\leq-0.57633$). With this selection we ended up with a sample of 8,809 anomalies, with 65.5\% them belonging to class Q. 

The set of anomalies is dominated by sources with noisy light curves (due in some cases to problems in the images that are not identified by the ZTF pipeline), and by sources with incorrect labels in the original catalogs (e.g., we found several variable stars classified as AGNs in the literature). Therefore, we highlight that our AD algorithm is useful not only to find intrinsic anomalous light curves, but also to detect problems in the datasets used for AGN variability analyses. 

We visually inspected the anomalies found by our AD algorithm and selected a sample of 75 promising CSAGN candidates. The full list of candidates is presented in Appendix \ref{appendix:cand_list}. The visual inspection is required, as a large fraction of our anomalies correspond to bogus candidates (like light curves with photometric issues, or incorrect labels reported in the literature). Further spectroscopic confirmation would be required to confirm the real nature of our sample.

A high fraction of our candidates present light curve shapes similar to those observed in X-ray light curves of X-ray binaries, which are generally associated to changes in the accretion state. The timescales of these variations in our set of candidates are of the order of hundreds of days, much smaller than the typical viscous timescales for accretion disks around SMBHs (order of thousands of years), and thus we need further analyses to discover the origin of these atypical variations. 

This work corresponds to our first attempt to search for CSAGN candidates in massive datasets. In the future, we plan to use this architecture, together with other techniques, to forecast the light curves of known type 1 and type 2 AGNs, and from this, detect variability behaviors that deviate from the normal variations expected for each source. For this aim, we will use data from the ZTF data releases, but also data from the ZTF alert stream, which should allow us to detect CSAGN candidates in real-time, and trigger follow-up to better pin down the nature  of this enigmatic phenomenon. 

\begin{acknowledgments}

This work was funded by:
project CORFO 10CEII-9157 Inria Chile (PSS, HL, LM, NSP);
ANID -- Millennium Science Initiative Program -- ICN12\_009 (PSS, JA, GCV, PE, LHG, AMA, FEB, FF); 
ANID, -- Millennium Science Initiative Program -- NCN19\_171 (AB); 
CATA-Basal - AFB-170002 (FEB);
the Competition for Research Regular Projects, year 2019, code LPR19-22, Universidad Tecnológica Metropolitana and the high-performance computing system of PIDi-UTEM (SCC-PIDi-UTEM— CONICYT—FONDEQUIP—EQM180180; JV);
ANID FONDECYT Postdoctorado Nº 3200250 (PSS); 
ANID FONDECYT Iniciaci\'on Nº 11191130 (GCV); 
ANID FONDECYT Regular Nº 1171678 (PE), 1190818 (FEB), 1200495 (FEB), and 1200710 (FF).

Based on observations obtained with the Samuel Oschin 48-inch Telescope at the Palomar Observatory as part of the Zwicky Transient Facility project. ZTF is supported by the National Science Foundation under Grant No. AST-1440341 and a collaboration including Caltech, IPAC, the Weizmann Institute for Science, the Oskar Klein Center at Stockholm University, the University of Maryland, the University of Washington, Deutsches Elektronen-Synchrotron and Humboldt University, Los Alamos National Laboratories, the TANGO Consortium of Taiwan, the University of Wisconsin at Milwaukee, and Lawrence Berkeley National Laboratories. Operations are conducted by COO, IPAC, and UW.

\end{acknowledgments}

\software{Google Colaboratory \citep{Bisong19}, Keras \citep{Keras}, Scikit-learn \citep{Pedregosa12}, and Tensorflow \citep{Tensorflow}. For graphical representations and data manipulation: Jupyter \citep{Kluyver2016jupyter}, Matplotlib \citep{matplotlib}, Numpy and Scipy \citep{numpy}, Pandas \citep{pandas}, Python \citep{van1995python}, and Seaborn \citep{seaborn}.}

\appendix

\section{List of promising CSAGN candidates}\label{appendix:cand_list}

Table \ref{table:cand} shows the coordinates, redshifts (reported by MILLIQUAS), previous classifications, and ZTF DR5 and alerts object IDs, of the 75 CSAGN candidates described in Section \ref{subsec:candidates}. The column ``plateau'' indicates whether the light curve of the source shows the plateau feature in its light curve. The column ``SDSS spectrum'' indicates whether the source has SDSS spectrum available. Sources with recent spectrum are marked with $\blacklozenge$ in this column. There are four candidates in this sample that were reported in the list of CSAGNs of \cite{Graham20} (objects with ZTF alert ID ZTF19aalmwdr, ZTF18aaxmbqw, ZTF18aahhdas, and  ZTF18aaqmmll). In addition, two candidates have been already confirmed spectroscopically as CSAGNs (objects with ZTF alert ID  ZTF18acbzncb and ZTF18aavsdfj;  M. Graham, private communication). These six confirmed candidates are marked with $\dagger$ in column ``ID ZTF alerts''. Finally, 28 candidates are identified as candidates from their CRTS light curves, and/or their ZTF DR3 light curves, using the techniques presented in \cite{Graham20} (Graham et al. in prep.), these sources are marked with $\star$ in column ``ID ZTF alerts''.

\begin{longtable}{cccccccc}
\caption{Selection of promising CSAGN candidates.}\label{table:cand}\\
\hline
\textbf{RA (deg)} & \textbf{DEC (deg)} & \textbf{redshift} & \textbf{class} & \textbf{ID ZTF DR5} & \textbf{ID ZTF alerts} & \textbf{plateau} & \textbf{SDSS spectrum}\\
\hline
\endfirsthead
\multicolumn{4}{c}%
{\tablename\ \thetable\ -- \textit{Continued from the previous page}} \\
\hline
\textbf{RA (deg)} & \textbf{DEC (deg)} & \textbf{redshift} & \textbf{class} & \textbf{ID ZTF DR5} & \textbf{ID ZTF alerts} & \textbf{plateau} & \textbf{SDSS spectrum}\\
\hline
\endhead
\hline \multicolumn{4}{l}{\textit{Continued on next page}} \\
\endfoot
\hline

\endlastfoot

34.47318&-19.972092&0.46&Q&299104400000216&ZTF19abkfpqk&Yes&No\\
66.427416&-10.535333&&X&354108100003201&ZTF19acnkupu&No&No\\
152.664006&-12.737412&0.42&type2&366103400000077&ZTF20aadswdl&No&No\\
35.812817&-5.317624&0.96&type2&401103200007922&ZTF19aadbbtz&No&No\\
151.046448&-3.613936&0.45&Q&417106400000034&ZTF19aagoauz$\star$ &No&No\\
221.653773&0.782385&0.72&QR&427114100001870&ZTF18acuszrd$\star$&No&Yes\\
229.643977&-1.090381&0.65&Q&428109200001687&ZTF19aascyqe$\star$&No&Yes\\
310.764753&-1.023893&1.19&Q&440111200002868&ZTF18abvvtxk&Yes&Yes\\
28.853979&2.471266&0.08&AR&450101200003599&ZTF18abqvdqu$\star$&No&No\\
34.740162&2.461193&1.73&Q&451102100000333&ZTF19abfqigm&Yes&Yes\\
31.842494&2.526327&0.66&Q&451103200005537& - &Yes&Yes$\blacklozenge$\\
38.714227&2.774584&1.17&Q&452107300008623&ZTF19abjgcof&Yes&Yes\\
129.477711&4.002007&0.56&Q&465107200003387&ZTF19aaufsep&No&Yes\\
143.825809&2.07099&0.64&QR&467103100002301&ZTF18aczejdj$\star$&Yes&Yes\\
174.278534&1.663312&0.19&Q&471101300000081&ZTF19aalmwdr$\dagger$&Yes&Yes\\
183.858915&3.842626&0.93&Q&473107200004410&ZTF18acvgvou$\star$&Yes&Yes\\
194.984234&3.409168&0.74&Q&474105300001176&ZTF20aavavif$\star$&Yes&Yes\\
194.992985&7.223641&1.5&Q&474113300001013&ZTF18acvwlvv$\star$&Yes&Yes\\
228.771345&4.170062&1.26&Q&479105200007871&ZTF19aailvld$\star$&Yes&Yes\\
224.519576&4.054482&0.69&Q&479108100002617&- &No&Yes\\
236.681566&2.138313&0.6&QR&481104200003568&ZTF19abbtqpb&Yes&Yes\\
336.760588&6.224071&1.23&Q&495111100005833&ZTF20abcxmfu$\star$&No&Yes$\blacklozenge$\\
333.860942&6.253339&0.22&Q&495112200000743&ZTF18aceiofb$\star$&Yes&No\\
193.365443&14.915562&0.25&Q&526115100009333&ZTF18aahhdas$\dagger$&Yes&Yes\\
206.858147&11.600358&0.84&Q&528107200000356&ZTF18ablqdss$\star$&Yes&Yes\\
310.077349&10.995846&0.08&Q&543108100012056&ZTF18aazmwqp&No&No\\
113.92998&20.215779&2.1&Q&566112100004425&ZTF18abwclfc&Yes&Yes$\blacklozenge$\\
144.242347&19.499183&0.38&Q&570111400005490&ZTF18accnmoe$\star$&No&Yes\\
202.796024&22.656575&0.96&Q&578114100000128&ZTF18aavsdfj$\dagger$&Yes&Yes\\
227.421616&18.766963&0.32&A&582108200000456&ZTF20aaznovd&Yes&Yes\\
228.182002&19.979198&0.8&Q&582112100014255&ZTF20aauwvfb&Yes&Yes\\
244.439728&16.548838&2.11&Q&584103100003077&ZTF19aailumj&Yes&Yes\\
333.436661&21.787462&0.65&Q&596114400000305&ZTF19acgfpvb&Yes&Yes\\
26.371489&28.705946&0.64&QR&603113300005168&ZTF19abdkgly&Yes&No\\
123.570806&25.493102&0.6&Q&616105200002737&ZTF18aagevkc$\star$&Yes&Yes\\
192.926178&24.076456&0.18&AR&626104200000260&ZTF18acahbqb$\star$&No&Yes\\
204.008817&23.63965&0.49&Q&627102200007292&ZTF19aailxfk&Yes&Yes\\
349.263518&29.416571&1.26&Q&646113100002570&ZTF20abwddhk&Yes&Yes$\blacklozenge$\\
4.257901&30.070426&1.3&Q&648101400002016&ZTF18abvztgk&Yes&Yes$\blacklozenge$\\
358.545135&35.605134&0.22&Q&648115300004761&ZTF18abmawjk&Yes&No\\
12.46894&33.53162&0.37&A&649109400004074&ZTF19abujzxs&Yes&No\\
12.624976&35.436805&1.56&Q&649113400004567&ZTF20abmdzgu&Yes&Yes$\blacklozenge$\\
47.684888&32.658078&0.12&A&654106200004523&ZTF19aacqijv$\star$&Yes&No\\
120.834624&29.763564&0.52&Q&663101300003892&ZTF18aagccsf&Yes&Yes\\
124.148074&33.796856&0.51&AR&664111300002498&ZTF20acfxnec$\star$&Yes&Yes\\
184.860032&32.023781&0.63&Q&672108300002204&ZTF19actnycp&No&No\\
185.403099&34.59464&0.29&A&672112100002964&ZTF18aaxmbqw$\dagger$&Yes&Yes\\
206.738679&36.547599&0.24&A&674113200005713&ZTF18aacdcxp&Yes&Yes$\blacklozenge$\\
218.187173&30.243154&0.35&QR&676103300001128&ZTF18aautsvy$\star$&No&Yes\\
215.7201&30.274046&2.2&Q&676104300000686&ZTF20aawwxkg&Yes&Yes\\
235.017706&35.847258&0.16&A&678115400001961&ZTF18aakhtca$\star$&Yes&Yes$\blacklozenge$\\
260.736568&32.240587&1.17&Q&681106400012690&ZTF18aayaxrf&No&Yes\\
348.418519&30.217598&0.86&Q&692101400001447&ZTF18abtclue$\star$&Yes&Yes\\
353.578057&32.08145&0.32&Q&693106300007466&ZTF18acbxoul&Yes&Yes$\blacklozenge$\\
134.863998&43.929443&0.35&A&709113200000994&ZTF20aayvktl&Yes&Yes\\
155.107655&41.561772&0.36&Q&712112100003746&ZTF18acbzncb$\dagger$&Yes&Yes$\blacklozenge$\\
180.15749&44.177165&1.37&QR&715116100000207&ZTF18aaqleut&Yes&Yes\\
192.080372&37.107331&1.52&Q&716103400002868&ZTF19adcfwam&Yes&Yes\\
188.569167&42.559662&1.54&Q&716116400003725&ZTF20aagxlkl&Yes&Yes\\
219.800624&37.085547&0.79&Q&719102400002793&ZTF19adcfvjv&Yes&Yes\\
220.51178&43.619104&0.23&Q&719114100002677&ZTF18aaqmmll$\dagger$&No&Yes$\blacklozenge$\\
237.316978&41.452011&0.93&Q&721110400013285&ZTF20aagiimu&Yes&Yes\\
243.249306&42.32787&0.23&QR&722111100000279&ZTF18aanlzzf$\star$&No&Yes\\
5.028384&45.133848&0.81&QR&736101200006573&ZTF20abcxodv$\star$&Yes&No\\
15.913292&47.5503&0.19&A&737105200000494&ZTF18abosuim$\star$&No&Yes\\
230.847407&49.377265&1.57&Q&760112100000758&ZTF18aczerbt$\star$&No&Yes\\
233.355747&50.192888&1.1&Q&760115400002305&ZTF20aalrlww&Yes&Yes\\
298.14922&49.97055&0.46&A&766113400010935&ZTF18aawvilc&Yes&No\\
177.603194&52.873391&0.7&QR&789103200005093&ZTF18aceyygm$\star$&No&Yes\\
220.21153&52.079437&0.31&A&793104100003683&ZTF19aaxezow$\star$&Yes&Yes\\
222.201328&53.340535&1.55&Q&793107300003189&ZTF20aapbkig&Yes&Yes\\
225.030254&56.60022&0.88&QR&793110200000811&ZTF19abzmptx&Yes&Yes\\
248.321942&54.399034&1.37&Q&795106100002111&ZTF19aawmmfv$\star$&Yes&Yes$\blacklozenge$\\
167.121585&64.858964&0.71&Q&818114300000932&ZTF19aaeduzf$\star$&Yes&Yes\\
257.828352&65.503293&1.09&Q&825115100005176&ZTF19aaqagri&Yes&Yes\\
\end{longtable}

\bibliography{bibliography.bib}
\bibliographystyle{aasjournal}



\end{document}